\newcommand\ha{{H$\alpha$}}
\newcommand\hb{{H$\beta$}}

\newcommand\LUM{\:{\rm erg\:s^{-1}}}
\newcommand\FLUX{\:{\rm erg\:cm^{-2}\:s^{-1}}}

\newcommand\VEL{\:{\rm km\:s^{-1}}}
\newcommand\DENS{\:{\rm cm^{-3}}}

\newcommand\OiL{[\ion{O}{1}] $\lambda 6300$}

\newcommand\OiiiL{[\ion{O}{3}] $\lambda\lambda 4959,5007$}
\newcommand\SiiL{[\ion{S}{2}] $\lambda\lambda 6717, 6731$}
\newcommand\NiiL{[\ion{N}{2}] $\lambda\lambda 6548, 6583$}

\newcommand\sii{[\ion{S}{2}]}
\newcommand\nii{[\ion{N}{2}]}
\newcommand\oi{[\ion{O}{1}]}

\newcommand\oiii{[\ion{O}{3}]}

\newcommand\hii{\ion{H}{2}}

\newcommand{\EXPU}[3]{\mbox{\rm $#1 \times 10^{#2} \rm\:#3$}}  
\newcommand{\POW}[2]{\mbox{$\rm10^{#1}\rm\:#2$}}
\newcommand{\SINGLET}[3]{[\ion{#1}{#2}]$\lambda #3$}
\newcommand{\MSOL}{\mbox{$\:M_{\sun}$}}

\documentclass[modern]{aastex61}
\shorttitle{SNRs in M33}
\shortauthors{Long et al.}

\usepackage{comment}
\usepackage[colorinlistoftodos]{todonotes}

\begin{document}

\title{MMT Spectroscopy of Supernova Remnant Candidates in M33}

\correspondingauthor{Knox S Long}
\email{long@stsci.edu}

\author[0000-0002-4134-864X]{Knox S. Long}
\affil{Space Telescope Science Institute,
3700 San Martin Drive,
Baltimore MD 21218, USA; long@stsci.edu}
\affil{Eureka Scientific, Inc.
2452 Delmer Street, Suite 100,
Oakland, CA 94602-3017}

\author[0000-0003-2379-6518]{William P. Blair}
\affiliation{The Henry A. Rowland Department of Physics and Astronomy, 
Johns Hopkins University, 3400 N. Charles Street, Baltimore, MD, 21218; 
wpb@pha.jhu.edu}

\author[0000-0002-0763-3885]{Dan Milisavljevic}
\affiliation{Department of Physics and Astronomy, Purdue University, 525 Northwestern Avenue, West Lafayette, IN 47907; dmilisav@purdue.edu}
\affiliation{Harvard-Smithsonian Center for Astrophysics, 60 Garden St., Cambridge, MA 02138; jraymond@cfa.harvard.edu}

\author[0000-0002-7868-1622]{John C. Raymond}
\affiliation{Harvard-Smithsonian Center for Astrophysics, 60 Garden St., Cambridge, MA 02138; jraymond@cfa.harvard.edu}

\author[0000-0001-6311-277X]{P. Frank Winkler}
\affiliation{Department of Physics, Middlebury College, Middlebury, VT, 05753; 
winkler@middlebury.edu}




\begin{abstract}

To date, over 220 emission nebulae in M33 have been identified as supernova remnants (SNRs) or SNR candidates, principally through \sii:\ha\ line ratios that are elevated compared to those in \hii\ regions.   In many cases, the determination of a high \sii:\ha\ line ratio was made using narrow-band interference filter images and has not been confirmed spectroscopically.  Here we present MMT 6.5\,m optical spectra that we use to measure \sii:\ha\ and other line ratios in an attempt to determine the nature of these suggested candidates.  Of the 197 objects in our sample,  120 have no previously published spectroscopic observations. We confirm that the majority of candidate SNRs have emission line ratios characteristic of SNRs. While no candidates show Doppler-broadened lines expected from young, ejecta-dominated SNRs ($\ga 1000$\,$\VEL$), a substantial number do exhibit lines that are broader than \hii\ regions.  We argue that the majority of the objects with high \sii:\ha\ line ratios ($>$0.4) are indeed SNRs, but at low surface brightness the distinction between H II regions and SNRs becomes less obvious, and  additional criteria, such as X-ray detection, are needed. We discuss the properties of the sample as a whole and compare it with similar samples in other nearby galaxies.

\end{abstract}




\section{Introduction} \label{sec:intro}

Although most Galactic SNRs were first identified as extended sources of non-thermal radio emission, most extragalactic SNRs have been identified through interference filter imaging as emission nebulae with  \SiiL:\ha\ line ratios that are elevated compared to those in \hii\ regions \citep[see, e.g.][and references therein]{long16}.  In \hii\ regions, particularly those of high surface brightness, the observed \sii:\ha\ ratios are typically of order 0.1 because most sulfur is photoionized to S$^{++}$. In contrast, most SNRs are observed to have \sii:\ha\ ratios of at least 0.4 because sulfur is found in a wide variety of ionization states in the extended recombination zone behind radiative shocks.  This expectation has support from a long series of radiative shock model calculations that generally confirm this expectation \citep[e.g.,][]{raymond79, hartigan87, dopita95, allen08}. 

The reason that most extragalactic SNRs have been identified optically reflects the history of  relative sensitivity of optical, radio, and X-ray searches for SNRs in external galaxies.  M33, at a distance of 817$\pm58$ kpc \citep{freedman01}, was one of the first galaxies where a search for SNRs was carried out.  The first three SNRs there were identified by \cite{dodorico78} and then confirmed spectroscopically by \cite{dopita80}.  Additional searches followed as instrumentation improved, in particular with the advent of CCDs \citep{long90,gordon98} so that by the turn of the century there were approximately 100 SNR candidates known in M33.  Most recently \cite[][hereafter Long10]{long10}  and \cite[][hereafter LL14]{lee14_m33} identified 137 and 199, respectively, partially overlapping sets of SNRs in M33, both from an examination of interference filter images obtained by \cite{massey06,massey07} as part a ground-based survey of Local Group galaxies.\footnote{Long10 also used a set of somewhat deeper but lower resolution images from the 0.6m Burrell Schmidt telescope, in addition to the LGGS.}  It is not surprising that these two sets of candidates are not identical; although both studies found the same {\em bright} SNR candidates,  at the faint end of the distribution, the criteria used to identify candidate SNRs become more subjective and confused. LL14 also surveyed extended regions in the northern and southern extremes of M33, beyond the region searched by Long10. 

The brightest SNRs are apparent in \sii\ and \ha\ images, but in order to search effectively for fainter remnants, continuum images must be subtracted from the emission-line ones.  The subtraction is never perfect, due to changing seeing conditions and color terms associated with different stars.  The initial selection of candidates is made by visual inspection, and extracting line ratios from the subtracted emission-line images is by no means straightforward.  Additionally, different investigators adopt different criteria for determining what is a {\it bona fide} candidate, such as requiring a certain morphology or not, excluding nebulae with interior blue stars or not, etc.  Furthermore, the pass bands for the \ha\ interference filters used by different observers is important, since virtually all such filters pass at least some of the \NiiL\ lines adjacent to \ha, ``contaminating" the \ha\ images by varying amounts and thus lowering the apparent \sii:\ha\ ratio. 
For all of these reasons and more, it is important to obtain spectra of as many of the SNR candidates as possible to clarify their status. As summarized by Long10, spectra of 85 of the 137 then-known SNRs or candidates had been obtained, with a variety of instrumentation.

In this paper, we describe a new spectroscopic study of the SNR candidates in M33.  Our goal was to obtain spectra of as many of the SNRs as possible with the same instrumental setup and to include as many as possible of the fainter SNR candidates, particularly those not observed previously, in order to obtain accurate \sii:\ha\ ratios and, in as much as possible, determine the status of the SNR candidates.  


\section{Observations and Data Reduction \label{sec:observations}}

\begin{figure}
\plotone{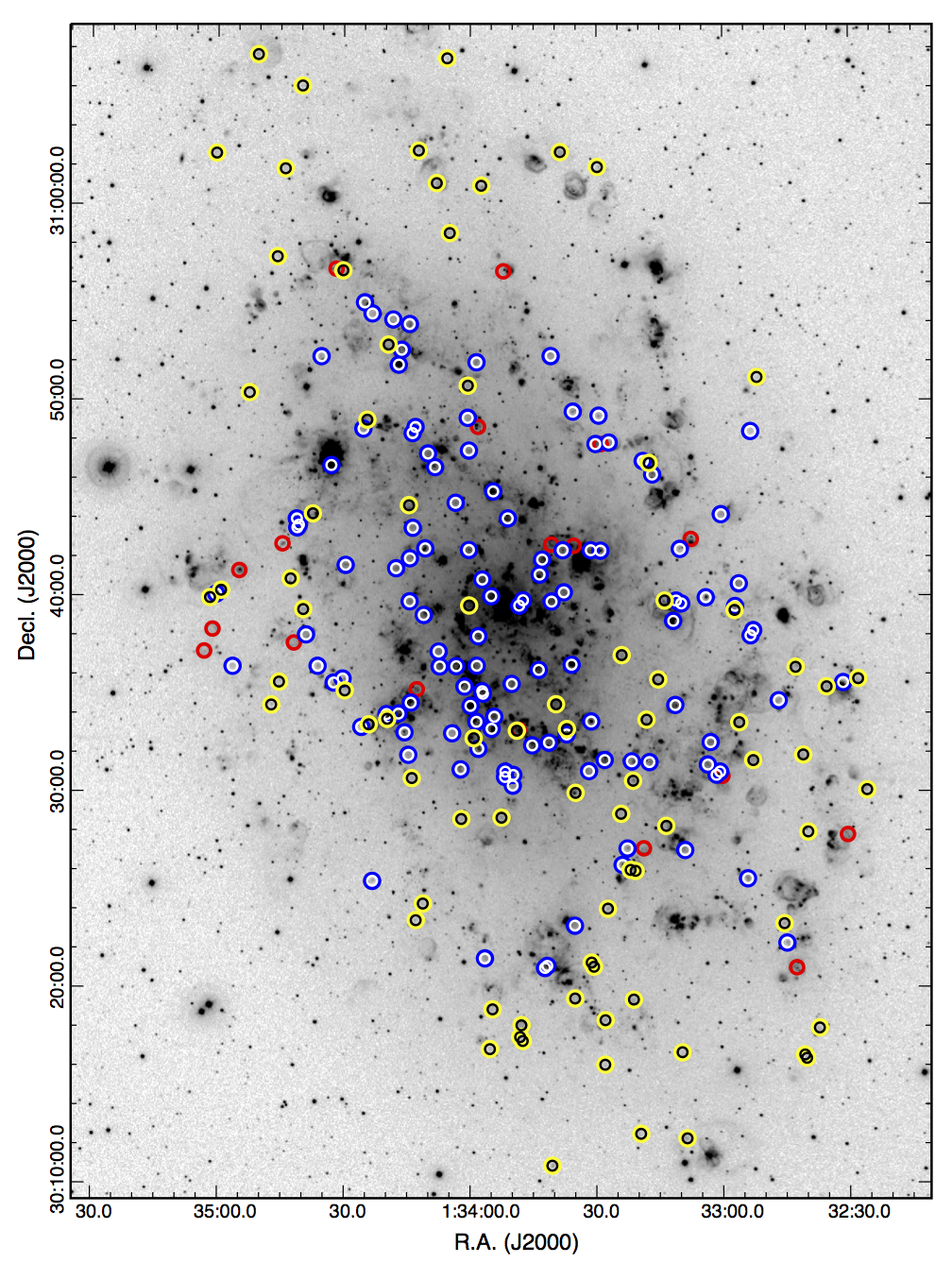}
\caption{An image of M33 from the 0.6m Burrell Schmidt telescope at KPNO, with the positions of the SNRs and SNR candidates indicated as follows:  blue/white circles indicate objects that appear in both the L10 and LL14 catalogs; red  circles are ones that appear only in L10; and yellow/black circles are ones that appear only in LL14. The field size is $44\arcmin \times 60\arcmin$, oriented north up, east left. 
\label{fig_overview} }
\end{figure}

In order to define our observing program, we first created a combined set of targets from the list of objects contained in Long10 and LL14. This was necessary because the LL14 list is not a simple superset of the list of Long10.  LL14 eliminated a number of sources from their consideration because they argued the sources were too large or not ``SNR-like" enough in  morphology to be considered SNRs.  They also remeasured source positions and concluded that some candidates were associated with a somewhat different set of filaments than Long10 had identified.  In identifying SNR positions we adopted the following (somewhat parochial) strategy of favoring positions from Long10.  We assumed that a SNR identified by LL14 was the same SNR as identified by Long10 if the position was within 3\arcsec\ of a Long10 SNR.  By this criterion, 119 of the 199 LL14 SNRs correspond to Long10 SNRs, and thus there are 80 LL14 SNRs that are not in the Long10 list.  There are also 18 Long10 SNRs not in the LL14 list.  Thus our initial sample comprised 217 objects, whose positions are shown on an \ha\ image of the galaxy in Fig.\ \ref{fig_overview}.
Recently, \cite{garofali17} have compiled their own list of SNRs from Long10 and LL14 and elsewhere.  Their list has 218 objects, three of which---XMM-081, XMM-089, and XMM-095---were not identified either by Long10 or LL14, and which based on our inspection of the LGGS images have no associated optical counterparts.  \citet{garofali17} also do not include two objects which we have listed as separate objects here: LL14-096 and LL14-174.

Some basic supporting data about the 217 objects for the purpose of this discussion are presented in Table \ref{table_snr_obs}.  This table contains (1) the source name we use for the object in this paper, (2,3) the position of the object, (4) the apparent size of the object in pc, (5) the galactocentric distance of the object in kpc, (6) the name of the object in Long10, (7) the name of the object in LL14, (8) whether the object has been detected in X-rays, (9) the reference to previous spectra of the the object, (10) whether we acquired a new spectrum, and (11) whether or not the spectra indicate that \sii:\ha\ ratio was greater than or equal to 0.4.  In creating the table, we have taken names and data from Long10 if the object existed in Long10 and used data from LL14 for the remainder.  We have not remeasured the apparent diameters of the objects.  For the purpose of this discussion, we have listed an object as X-ray detected if the significance of the detection was 3$\sigma$ with {\em Chandra} by Long10 or with XMM-{\em Newton} by \cite{garofali17}.  We record an object as \sii:\ha\ confirmed if we find a \sii:\ha\ ratio greater than or equal to 0.4, or if it was not observed by us, but previous observers have reported a similarly high ratio.  Objects with no entry in this column are objects without (to the best of our knowledge) any previous spectroscopic followup.

Our new spectroscopic observations were carried out with Hectospec \citep{fabricant05}, a moderate resolution ($\sim 5$ \AA) multi-fiber spectrograph available on the 6.5m MMT.  We used the 270 lines mm$^{-1}$ grating, which gives a total spectral coverage
of 3600 to 9100 \AA\@.  Each fiber has a core diameter on the sky of 1.5$^{\prime\prime}$, spanning $\approx 6$ pc at the 817 kpc distance of M33. 

The observations took place over eight nights in 2016 October and 2016 November under good conditions.  We obtained spectra of 197 out of the list of 217 objects:  110 objects in both lists, 18 objects in the Long10-only list, and 69 objects in the LL14-only list.  For comparison purposes, we also obtained spectra of 23 bright \hii\ regions, but since a far larger spectroscopic survey of \hii\ regions in M33 has recently been reported by \cite[][hereafter Lin17]{lin17}, we have  used our own  \hii\ region spectra only for estimating instrumental line widths. Individual exposures of 1800 s were obtained and combined to create total exposures that varied from $\sim$ 2 to 10 hr per target.

\begin{figure}
\plotone{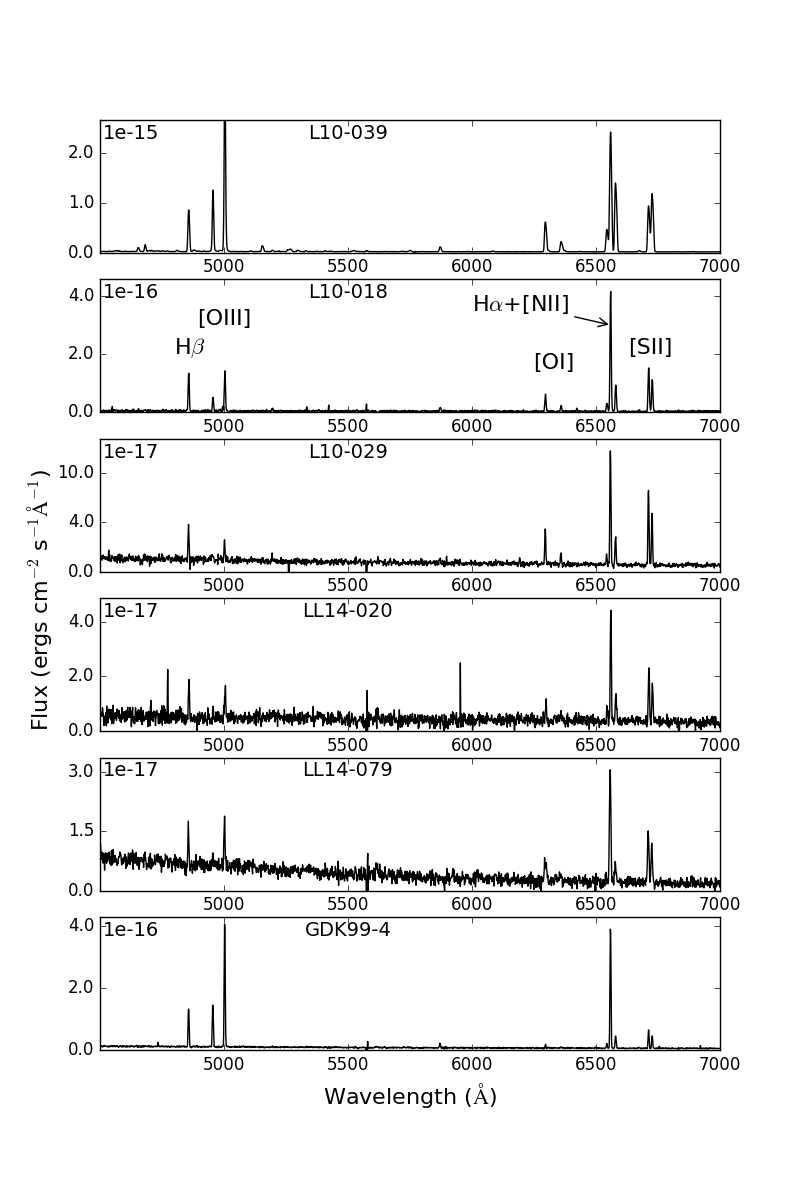}
\caption{Example spectra. The spectra of  5 SNR candidates are shown in the upper 5 panels of the figure.  The various spectra were selected  to indicate the quality of the data as a function of surface brightness.  The continuum in the spectrum of L14-079 arises from the fact that in addition to emission from the nebula, the fiber captured light from stars along the line of sight as well. The spectrum of a moderately bright \hii\  region is shown in the bottom panel, for comparison.  The lines of interest are labeled in the second panel.  The flux scaling numbers in the upper left corner of each panel should be applied to the values on the vertical axis of that panel. \label{fig_spectra} }
\end{figure}

All of the data were reduced using the standard Hectospec pipeline (HSRed Version 2.0)\footnote{https://www.mmto.org/node/536}, which applied wavelength calibration, performed cosmic-ray rejection, subtracted sky emission using $>20$ sky fibers that were averaged and scaled in intensity, and corrected for telluric absorption features. For flux calibration, we used multiple observations of the spectrophotometric standard star BD+28-4211.  Many of the objects were observed multiple times.  Since the spectra of individual objects observed over multiple nights were similar, we combined the spectra for various nights weighting the spectra by exposure time on each night. In Fig.\ \ref{fig_spectra} we show several of our SNR candidate spectra, selected to indicate the typical quality of the spectra and sky subtraction for objects with a range of brightness.  For comparison, we also show one of our \hii\  region spectra.



We extracted line fluxes for several important  emission lines expected in SNRs: \hb, \OiiiL, \OiL, \ha, \NiiL\ and \SiiL, using the same Gaussian fitting routine we have used for extracting line fluxes from 1-D spectra on other projects \citep [e.g., M83, see][]{winkler17}.  The results of these fits are summarized for the SNR candidates in Table \ref{table_snr_spec}.  The columns in the table include  the object name,  the \ha\ flux\footnote{This is the flux sampled by a 1\farcs 5 diameter fiber, {\it not} the total flux from the entire object.  Since all the objects are larger than the fibers that sampled them, this flux is best interpreted as a surface brightness.} in units of \POW{-17}{\FLUX}, and the fluxes for other lines relative to \ha.  
For the doublets where the line ratio is constrained by atomic physics we  list only the stronger of the lines.  For \ion{S}{2}, where the ratio of the two lines in the doublet is density-sensitive, we list the ratio between  the \SINGLET{S}{2}{6717} and \SINGLET{S}{2}{6731} lines, as well as their sum relative to \ha.  We follow the convention where the \ha\ flux is taken to be 300, except for the ratio of the two \sii\ lines where we give the simple ratio between those two lines.  We have not quoted errors for the various values contained in the table, as it is unclear how to do this in a robust manner.  Based on a comparison of the flux ratios in spectra of the same object observed on multiple nights, the ratio errors for bright, well-observed lines are typically less than 15\%.  Lines where the accuracy of the line ratio is clearly worse than this have entries preceded by a $\sim$ symbol.  



\begin{figure}
\plotone{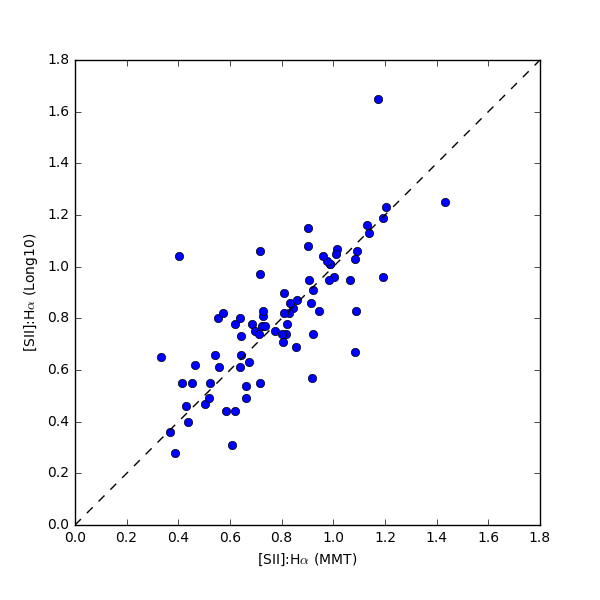}
\caption{The \sii:\ha\ ratio of SNRs measured from spectra obtained here with the MMT compared to  \sii:\ha\ ratios from spectra of SNRs collated by Long10.  Ideally all the points would fall along the dashed 1:1 line.  Given the the variety of instrumental setups involved, the agreement is probably as good as one could expect. \label{fig_old_new} }
\end{figure}

Based on the tabulation provided by Long10, there are 85 SNRs in M33 that already had spectra;  of these, we have new spectra of 77.  A comparison of the \sii:\ha\ ratios from  earlier measurements with those from this paper is shown in Fig,\ \ref{fig_old_new}.  Given  the variety of instruments and different slits and fibers used in the earlier estimates, and the fact that none of the apertures or fibers covered an entire SNR, there is quite good agreement between the past spectra and the new measurements.

There are eight  objects for which historical spectra exist and for which we did not get new  spectra: 
L10-017, 
L10-021, 
L10-047,
L10-056,
L10-083,
L10-095,
L10-102, and
L10-129.
In the earlier spectra, all of these objects have \sii:\ha\ ratios $>0.4$ and thus they should be considered valid SNRs based on this criterion.  As such we have listed them as ``\sii:\ha\ Confirmed" in Table \ref{table_snr_obs}.


\section{Results}


As noted in Section 1, the primary criterion used to discriminate optical SNRs from photoionized nebulae is that the ratio of \sii:\ha\ be $\gtrsim 0.4$. This criterion has been effectively applied for identifying objects in many galaxies, though the reliability depends on a number of variables including how low in surface brightness one surveys, how the metallicity varies, and various characteristics of the \hii\ regions.  At some stage, questions of size and morphology must also be taken into account. For brighter nebulae in most nearby galaxies, the \sii:\ha\ criterion gives a clean separation between SNRs and \hii\ regions.  But for some galaxies, and especially for low surface brightness objects, many nebulae have \sii:\ha\ ratios in the range 0.2 -- 0.5 that blur the dividing line between SNRs and \hii\ regions.  Of course observational uncertainty also becomes increasingly significant for fainter objects, further obfuscating the demarcation.

\begin{figure}
\plotone{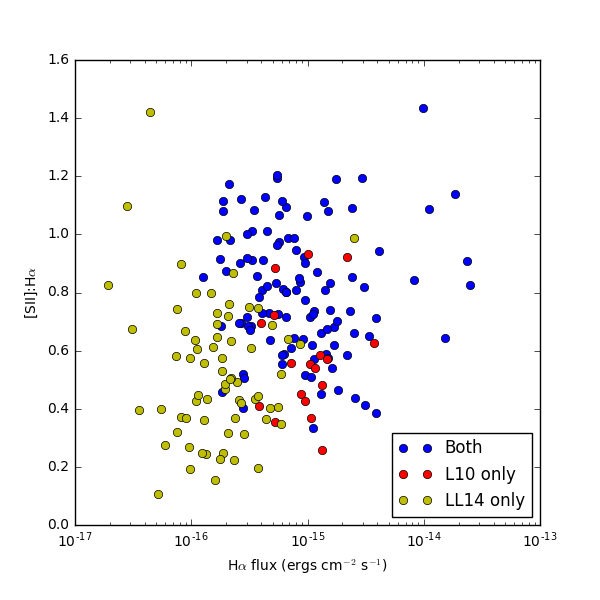}
\caption{The \sii:\ha\ ratio of SNRs measured from spectra obtained with the MMT as a function of the \ha\ flux obtained from the spectra.  The SNR candidates that appear in both Long10 and LL14 are plotted in blue; those from Long10 only in red, and those from LL14 only in yellow. The LL14-only candidates are generally fainter than the others, and have a higher disperion in the \sii:\ha\ ratio.  \label{fig_ha_ratio} }
\end{figure}

Our original candidate sample comprised 217 sources, of which we obtained spectra for 197.  The \sii:\ha\ ratios for these 197 are plotted as a function of the measured \ha\ flux  in Fig.\ \ref{fig_ha_ratio}.  Of these, 170  (86\%) satisfy the 
\sii:\ha $ > 0.4$ criterion.  Further breaking these down, 108 of the 110 SNR candidates that appear in {\em both} the Long10 and LL14 lists meet the criterion, as do 15 of 18 identified only  by  Long10, but only 47 of 69 of those identified only by LL14.

It is perhaps not surprising that a higher percentage of the LL14-only objects have lower ratios; these are systematically fainter than the other candidates.  We have inspected many of the LL14 objects on the LGGS images, and while those of  brightness comparable to the fainter Long10 objects look like reasonable candidates, many of the faintest LL14 objects are exceedingly faint in \ha\ and are barely visible at all in \sii. Now that we have spectra of many of these objects, we find that a number have low \sii:\ha\ ratios, even given the uncertainties that attend the relatively low signal-to-noise spectra.  Our spectra, together with their morphology on LGGS images, lead us to reject the following LL14 objects as bona fide SNR candidates going forward:  LL14-004, 009, 014, 032, 046, 048, 057, 059, 109, 133, 134, 188, and 198. Several additional objects should be considered marginal SNR candidates at best, given observational uncertainties in their determined ratios, and given the assessment of M33 \hii\ regions in the next section. However, we retain all the remaining  Long10 and LL14 SNR candidates in the color-coded plots (such as Fig.\ \ref{fig_lin}) presented below.



\begin{figure}
\plotone{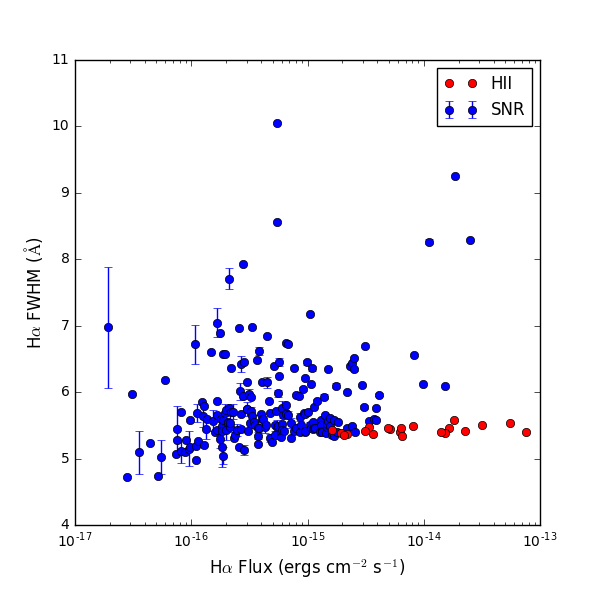}
\caption{In fitting the lines, we fit a single FWHM for the 3 lines \ha\ + \nii\ complex.  The figure shows the FWHM for the line complex as a function of \ha\ flux.    The FWHM for \hii\  regions all cluster around 5.4, whereas the SNRs have a broader distribution that is asymmetric toward higher values.  The formal errors for a subsample of the values are shown. Plots made with of the \sii\ line complex are very similar. \label{fig_vel} }
\end{figure}

In addition to measuring the line fluxes for the objects in our sample, we also measured the line widths (full width at half maximum; FWHM).   In fitting the lines, we fit a single width to closely spaced line complexes, viz., a single FWHM for \hb\ and \oiii, one for \ha\ and \nii, and one for the \sii\ doublet.  A comparison of the velocity widths for all three complexes shows very similar results; hence, we concentrate on the \ha-\nii\ fits, where the signal-to-noise is highest. In Fig.\  \ref{fig_vel} we show the FWHM for both SNRs and \hii\ regions, plotted as a function of \ha\ flux.   For the \hii\ regions, the fits are tightly clustered about the mean width of 5.43 \AA\ with an RMS dispersion of only 0.06 \AA\@.  But the FWHM distribution for the SNRs has a mean of 5.85 \AA\  with dispersion 0.78 \AA\@.  As expected, the dispersion is larger for fainter objects with lower S/N, but it is clear that the FWHM distribution for the SNRs skews to higher than instrumental values.  

One expects that the material in \hii\ regions to be simply that of the ambient ISM. Hence, their emission line widths should be essentially the instrumental value.  For SNRs, the lines will be Doppler broadened with a velocity characteristic of their shock velocity.\footnote{The thermal broadening for H is only about 20 $\VEL$, too low to be measured at our spectral resolution.  For heavier elements, the thermal broadening is
considerably less.   But since the bulk velocity of material behind a radiative shock is very close to that of the shock
and since a typical line of sight through the SNR includes multiple regions on both sides of the SNR, the observed 
line broadening should be of the same order as the shock speed in the denser gas encountered by the SNR shock.  Because optical radiation in most SNRs arises from radiative secondary shocks propagating into denser gas, the observed broadening will be less than that of the primary SNR shock.}  
If one assumes that the observed resolution
\begin{equation}
FWHM_{obs}=\sqrt{FWHM_{inst}^2+FWHM_{SNR}^2 }
\end{equation}
then at \ha, a width of 5.85 \AA\ corresponds to velocity broadening of 100$\VEL$, typical of older SNRs.   A FWHM of 7 \AA\ corresponds to 200$\VEL$, and 10 \AA\ would correspond to 380$\VEL$, neither of which is unusual for radiative shocks. 
However, to the extent that we have chosen the brightest sections of the SNR to observe, we have chosen limb-brightened regions that are seen edge-on, biasing the FWHM to values below the shock speed.
We have found no obvious correlations between the FWHM and other properties of the SNRs, such as diameter.
And in particular, we find {\it no} objects with velocities approaching 1000 $\VEL$ or more, such as one finds in young, ejecta-dominated SNRs like Cas A \citep{Milisavljevic12}. Furthermore, we see no evidence for wildly discrepant chemical abundances as seen in ejecta-dominated SNRs.  While this is perhaps not surprising for a sample assembled based on the \sii:\ha\ ratio, in Long10 we also examined the LGGS \oiii\ images and were not able to identify any oxygen-rich candidate SNRs.

\subsection{Reddening}

\begin{figure}
\plotone{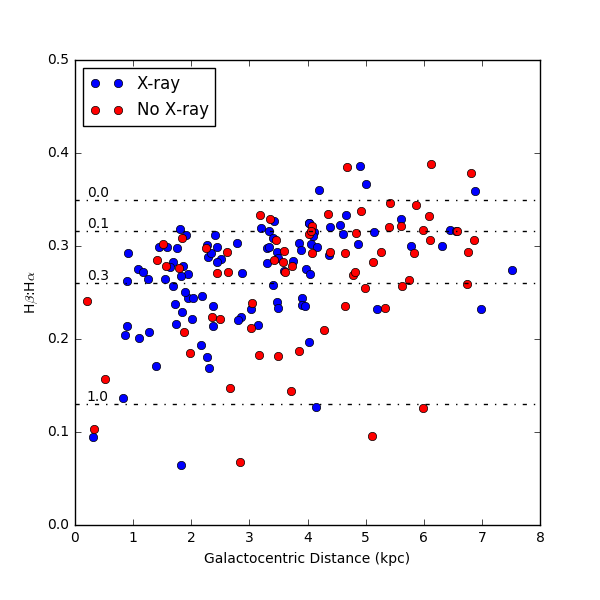}
\caption{The \hb:\ha\ flux ratio as a function of galactocentric distance. The dashed lines indicate reddening values (from top to bottom) of E(B-V) = 0.0, 0.1, 0.3 and 1.0. \label{fig_reddening} }
\end{figure}

The intrinsic value of \hb:\ha\ at low optical depth is constrained by the physics of recombination to be 0.35 \citep[e.g.,][]{osterbrock06}, but absorption due to, or more correctly scattering by, dust grains along the line of sight leads to lower observed ratios.    The observed \hb:\ha\ ratio for the SNR sample for which we have spectra is shown in Fig.\  \ref{fig_reddening}, as a function of galactocentric distance.  A very similar figure could have been made for the Lin17 \hii\ regions. There is a large variation in \hb:\ha, reflecting at least in part the location of candidates above or below the galactic plane of M33 as well as local variations in the amount of dust.  There is also an overall trend of decreasing absorption with larger galactocentric distance, due to the generally lower density (for both gas and dust) farther from the center of the galaxy.  The X-ray-detected SNRs do not exhibit significantly less reddening than the non-X-ray-detected ones, presumably because the overall reddening is still fairly low.  A reddening of $E(B-V)\approx 0.3$, typical of the value seen in the inner part of the galaxy, corresponds to an effective hydrogen column density of \EXPU{1.7}{21}{cm^{-2}}.

\subsection{Density Effects}

\begin{figure}
\plottwo{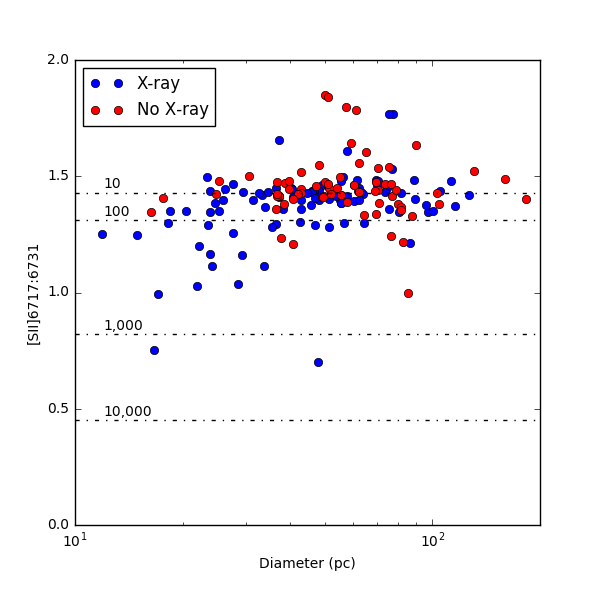}{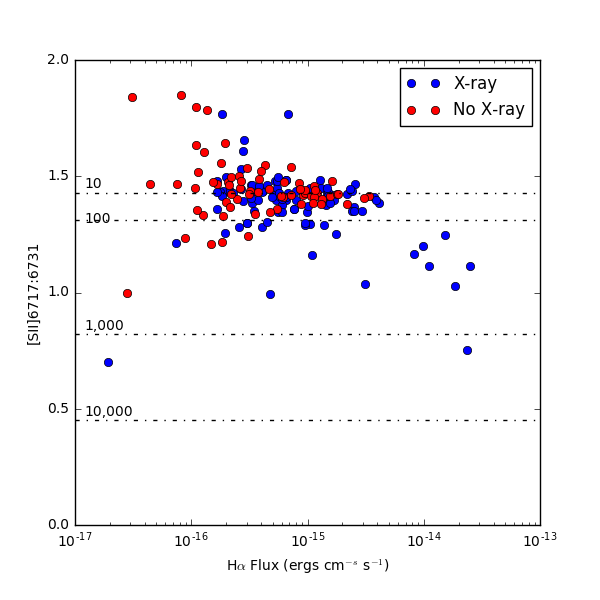}
\caption{The density-sensitive line ratio \sii\,$\lambda$6717:\sii\,$\lambda$6731 plotted as a function of SNR diameter (left panel) and flux (as measured through a 1.5\arcsec\ diameter fiber, right panel).  Only objects with \sii:\ha\ greater than 0.4 have been plotted.
Objects with X-ray detections and those without are plotted separately.  
The dashed lines indicate electron densities of 10, 100, 1,000 and 10,000 cm$^{-3}$ from top to bottom, respectively \citep{osterbrock06}.  
Most SNRs have  ratios near the low density limit.    Those that show higher densities lie preferentially at smaller diameters and higher flux levels.
\label{fig_density} }
\end{figure}

The ratio of \sii\ 6717:6731 is a well-known density diagnostic, driven by the importance of collisional excitation/deexcitation of the upper levels \citep[e.g.,][]{osterbrock06}.  This line ratio is presented in Fig.\ \ref{fig_density} for the 170 SNR candidates with a \sii:\ha\ ratio greater than 0.4, plotted as a function of SNR diameter (left panel) and flux through the 1.5\arcsec\ fibers (right panel).  The 112 candidates with X-ray detections are plotted separately from those without X-rays.  The ratios for most of the objects indicate that optical emission arises from material with density $n < 100 ~ \DENS$.  The left panel of the figure shows that at diameters $\lesssim 40$ pc, the fraction of SNRs detected in X-rays, and the fraction with higher density, are both greater than for SNRs with larger diameter. 
The right panel shows that, with considerable scatter, SNRs with higher flux tend to be detected in X-rays and also to show higher densities.  The trends with diameter are also seen for SNRs in M83 \citep{winkler17}.  
 The \ha\ emission per unit area of a shock
is proportional to the pre-shock density, and it increases with shock
speed at a rate between $V_s$ and $V_s^2$.  The density in the 
region that emits [S II] increases with the ram pressure, $n_0 V_s^2$,
though the relation is complicated by the contribution of magnetic
pressure.  Therefore, one expects the density derived from [S II] to
increase with the \ha\ flux, and that is borne out in the figure.  It is not entirely clear which of the two effects, density or shock velocity, dominates, though the absence of any truly high velocity SNRs  and the fact that soft X-ray emissivity (and hence detectability) peaks just as SNRs are entering the radiative phase favors density as the primary cause.   On the other hand, if one compares the ratio of \sii\ 6717:6731 for candidates with \sii:\ha\/ greater than 0.4 and measured line widths less than or greater than 5.85 \AA\ (corresponding to velocity width of 100 $\VEL$),  one finds the \sii\ 6717:6731 ratio is 1.42$\pm$0.13 for the ``narrow line'' objects' and 1.28$\pm$0.23 for the ``broad line'' objects, suggesting that shock velocity does play a role.
This is an area where systematic studies at higher spectral resolution  of a large number of SNRs in a galaxy like M33 would help.



\section{Discussion}
\subsection{Confirming Bona Fide SNRs: Validity of the \sii:\ha\ Ratio}

\begin{figure}
\plottwo{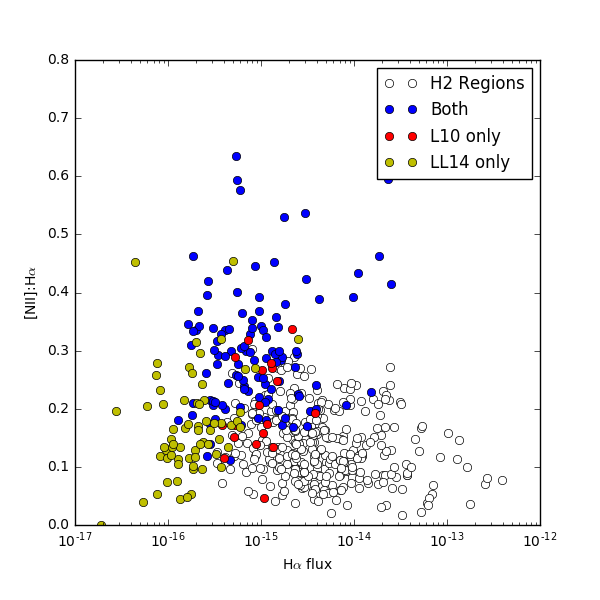}{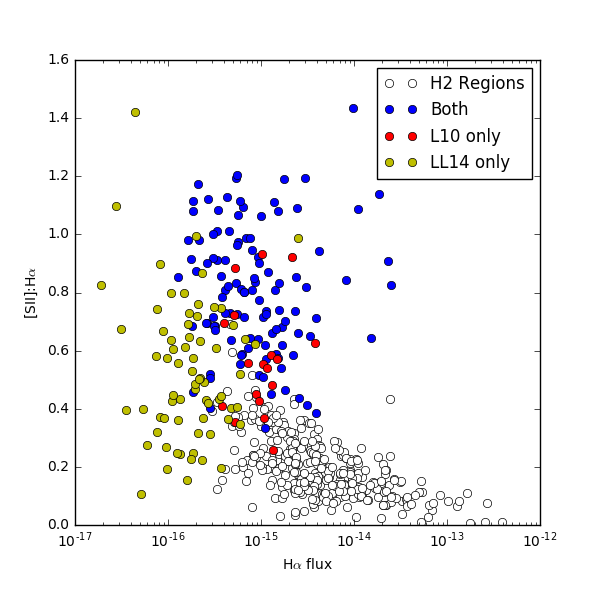}
\caption{Comparison of the \nii:\ha\ (left) and \sii:\ha\ (right) ratios of SNR candidates to the Lin17 \hii\ regions after pruning the Lin sample as discussed in the text.  The SNR sample is split into objects reported in both L10 and LL14, those in L10 only, and those in LL14 only. The ratios are plotted as a function of \ha\ flux, which is essentially a surface brightness through the filled Hectospec fibers.  \label{fig_lin}} 
\end{figure}

There are important caveats when using the \sii:\ha\ ratio to distinguish  \hii\ regions from SNRs. Perhaps most significant is that not all photoionized gas displays a low \sii:\ha\ ratio at faint surface brightness.  Diffuse ionized gas (DIG) is known to be photoionized and yet can have a high \sii:\ha\ ratio, due to the low density of the gas and its distance relatively far from the ionizing sources \citep{reynolds85,wood10}. The same effect is seen on the rims of large \hii\ complexes, which often have regions with elevated \sii:\ha\ ratio compared with the interior. \cite{blair97} encountered this situation when investigating the Sculptor group spirals NGC 300 and NGC 7793, where the \sii:\ha\ criterion broke down as a clean diagnostic.  Also, at ground-based spatial resolution, a diffuse patch of DIG may not be clearly distinguishable from a faint SNR, so a morphology criterion does not necessarily clarify the object identifications. Many of the SNRs and candidates in M33 have substantially lower surface brightness than the prominent \hii\ regions (Fig.\ \ref{fig_lin}, left).  Thus,  confusion near the \sii:\ha\ = 0.4 criterion (or any particular fixed value) used to separate \hii\ regions from SNRs can be a concern, especially for fainter objects.

Fortunately, Lin17 have recently conducted an extensive spectroscopic survey of \hii\ regions in M33  using  instrumentation identical to our own study.
We can use their results  to investigate the full range of observed ratios in a broad sample of emission regions and not be limited to the brightest \hii\ regions that have typically been observed previously. Lin17 reported spectra of 413 positions in M33, distributed over a large range of surface brightness and galactocentric distance.  Starting from a fairly low-resolution star-subtracted \ha\ image of M33 \citep[taken from the 0.6m Burrell Schmidt telescope, with a resolution of $\sim 4\arcsec$, as described by][]{hoopes01}, they selected fiber positions quasi-automatically using an algorithm that identified extended peaks of \ha\ emission. To avoid confusion with poorly subtracted stars, Lin17 avoided positions coincident with cataloged 2MASS sources,\footnote{Nevertheless, a number of \cite{lin17} spectra still include optically-bright stars.} but they appear not to have applied any other criteria to their target selection.  Consequently, a number of the Long10 and LL14 SNR candidates were incidentally included in their survey.  We used the Lin17 positions to create a region file that we overlaid on M33 images from the LGGS.  Through visual inspection of the images with Long10 and LL14 candidates also outlined, we eliminated 35 of the Lin17 spectra that were actually coincident with or overlapped SNR candidates.


In reviewing the other Lin17 fiber positions on the star-subtracted \ha\ images, we elected to eliminate spectra where the Lin17 algorithm had placed fibers on faint arcs or rims of very large, diffuse structures, or where there were multiple fibers placed on a single large diffuse nebula. We also eliminated a few fiber positions that appeared to lie on regions where no clearly-defined structure appeared to be present in the LGGS images. This resulted in the removal of another 38 fiber positions from consideration.  


The Lin17 fiber positions (not including the 73 rejected) sample a diverse set of emission nebulae when projected on the LGGS images, which have higher spatial resolution than those used by Lin17. Some are on bright, well-defined regions; some are on bright regions but displaced from the peak emission; others are at various locations in large \hii\ region complexes; while still others are located on faint, diffuse patches that are nevertheless discernible on LGGS images.  We suspect that some of the latter could be DIG, but have adopted the entire remaining set of 340 spectra as our ``\hii\ region" comparison sample in the figures that follow.


In Fig.\ \ref{fig_lin} we show the \sii:\ha\ ratios for our sample of SNR candidates, and the pared-down Lin17 sample of \hii\ regions, as a function of the \ha\ flux as observed through the 1.5\arcsec\ Hectospec fibers.  For  $F_{{\rm H}\alpha} \gtrsim 2 \times 10^{-15} \FLUX$, there is a relatively clean separation between the SNRs and \hii\ regions.  As seen in many previous studies, all of these brighter SNR candidates have \sii:\ha\ $> 0.4$, while the ratio is significantly lower for almost all of the brighter \hii\ regions.  At $F_{{\rm H}\alpha} \sim 10^{-15} \FLUX$ and below, however, the distinction becomes increasingly blurred.  In general, these fainter ``objects'' (in both samples) are less well-defined than the brighter ones.  Even if they are distinct objects, instrumental effects such as the difficulty in subtracting truly diffuse \ha\ emission that may overlie the object, and lower signal-to-noise in the spectra,  contribute to the dispersion in observed  \sii:\ha\ ratios for these faint objects.  

Ambiguity between SNRs and \hii\ regions can, to some degree, be attributed to advancing technology.  Early studies of M33 \citep[e.g.,][]{dodorico78, blair85} as well as elsewhere \citep[e.g.][]{levenson95} found clear distinctions based on the \sii:\ha\ ratio, but these studies were limited to the brightest and best defined objects.  Subsequent imaging and spectroscopy of increased depth and sensitivity have led to vastly larger samples that encompass objects that are an order of magnitude or more fainter than previously achievable.  Consequently, inclusion of fainter objects has also led to many for which the identification as photoionized versus shock-heated is less clear without confirmation from other criteria, such as radio or X-ray emission or secondary indicators in the optical spectra.




The MMT data for M33 clearly indicate that bright objects with high \sii:\ha\ ratios are a fairly distinct set of objects from the bright \hii\ region population;  the long-standing comparison of such spectra with predictions from shock models  provides strong evidence for shock heating.  However, at lower surface brightnesses the two distributions begin to merge.  Hence, we should expect that there is some contamination of the SNR candidate catalog by \hii\ regions, and equally likely that the SNR candidate list may be incomplete at these fainter levels.  

Along these lines, about half of the LL14-only candidates in Fig.\ \ref{fig_lin} with \sii:\ha\ ratio $< 0.4$ and low \ha\ flux are unlikely to be SNR candidates based on their morphology.  These objects appear to form an extension of the \hii\ region points to even lower surface brightnesses.  However, a number of the LL14-only objects, including some of the faintest, clearly have elevated \sii:\ha\ ratios consistent with their being shock-heated.  Such confusion is not limited to LL14-only candidates.  Fainter Long10 candidates are potentially uncertain as well, with ratio values that overlap those of the \hii\ region sample at comparable flux.



\subsection{Other Optical Diagnostics}

In principle, secondary criteria such as object morphology in the imagery or other supporting line ratios could be considered for uncertain objects in an attempt to clarify their nature.  In developing the sample for Long10 for example, we required spatially-extended candidates to show clear evidence of at least a partial shell structure in the LGGS images.  

Among additional emission line ratios, the \oi\ $\lambda$6300:\ha\ ratio is a possible  criterion, and in certain cases it may be a useful discriminant.  Shock-heated gas often shows elevated \oi\ emission in its cooling tail, just as it shows elevated \sii, and it is not normally observed  in even low-ionization photoionized regions because the normal UV background in the ISM can keep it ionized.  Observationally, measuring the \oi\ lines spectroscopically is complicated for objects in galaxies with near zero red (or blue) shift, because of the difficulty in subtracting night sky emission.  In our recent study of SNRs in M83 \citep{winkler17}, the redshift of $513 \VEL$ is sufficient that we were able to measure \oi\ emission in 110 of 118 SNR candidates. Resolving \oi\ emission lines in M33, with a blueshift of $179\VEL$, from the night sky lines is more challenging.
Nevertheless, we have measured clear \oi\ emission in the spectra of 93 of the 197 SNR candidates for which we have spectra.  Detection of \oi\ emission separate from the night sky is most difficult for the faintest objects with low signal-to-noise, which unfortunately is where it would be most valuable as a discriminant for the confused objects. 

\begin{figure}
\plotone{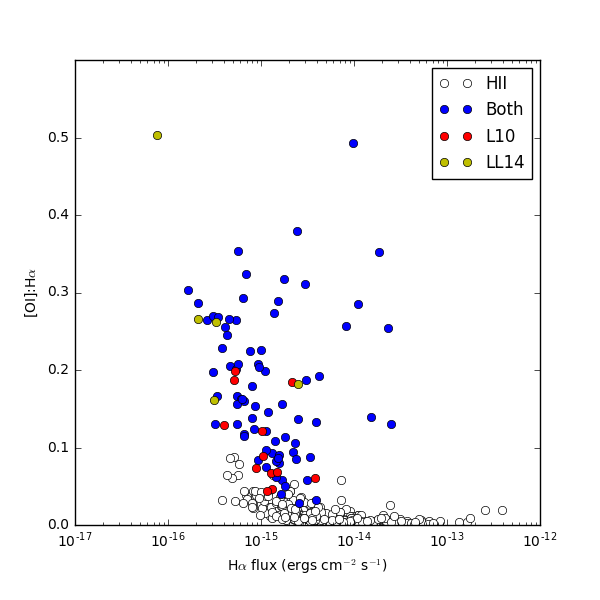}
\caption{Comparison of the \oi:\ha\ ratios of SNR candidates to the Lin17 \hii\ regions (after pruning the Lin sample as discussed in the text), as a function of \ha\ flux. The SNR sample is split into objects reported in both L10 and LL14, those in L10 only, and those in L14 only. 
\label{fig_o1} }
\end{figure}

In Fig.\ \ref{fig_o1} we show the \oi:\ha\ ratio for those objects where we were able to obtain reliable fits to the spectra, as well as for all of the pruned sample of Lin17 objects for which they give \oi\ fluxes.  \oi\ is generally far stronger for SNRs than for \hii\ regions, but there are a few SNRs for which \oi:\ha $\lesssim 0.05$, as well as a few \hii\ regions with higher ratios.  As expected, this is especially true at low surface brightness.  As with the \sii:\ha\ ratio, SNRs generally have significantly higher values than \hii\ regions, but there is no bright line that can cleanly separate the two classes of objects.  
Hence, rather than try to make a determination of object type for the objects at the lowest surface brightnesses, we simply point out the confusion.  At low surface brightness in particular, it is especially important to obtain confirming information from another wavelength band before declaring a SNR candidate to be a bona fide SNR.

Another possible criterion for distinguishing SNRs from \hii\ regions is the velocity broadening of the emission lines.  Lin17 do not give linewidths for their large sample of \hii\ regions, but as noted in Sec.\ 3, the 23 bright \hii\ regions that we observed all have line widths that are consistent with that of the Hectospec instrument configuration.  Our SNR candidates have line widths that are measurably greater in many cases, but for the fainter ones the spectral resolution and S/N are not high enough to measure the broadening from SNR shocks with velocity $\lesssim 100 \VEL$, as are found in most old SNRs.  

Much as we might like to identify a ``gold standard'' sample of SNRs identified based on their optical properties alone, we have not done this, since it would undoubtedly exclude many actual SNRs (especially fainter ones).

\subsection{X-ray and Radio Diagnostics}

At X-ray wavelengths, SNRs are extended sources with soft (line-dominated) X-ray spectra compared to most X-ray binaries of comparable luminosity (and the many background AGN that contaminate X-ray catalogs of nearby galaxies).  \hii\ regions are also extended X-ray sources, but they are, with the exception of a few giant \hii\ regions such as NGC\,604 in M33 \citep{tullmann08}, very much fainter than SNRs.  Because M33 is smaller in angular size than M31, is less inclined, and has less foreground absorption,  M33 is perhaps the best studied spiral galaxy at X-ray wavelengths.  Very sensitive X-ray surveys of M33 have been made both with {\em Chandra} \citep{tuellmann11} and XMM-{\em Newton} \citep{{williams15},{garofali17}}.  {\em Chandra}'s exquisite X-ray optics made it possible to image a number of known SNRs in M33, and to perform a search for new ones (though few new SNRs were found).  The number of new SNRs discovered in X-rays is small because (a) optical CCD technology enabled sensitive searches for extragalactic SNRs well before X-ray technology began to catch up, and (b)  most SNRs in even nearby galaxies are faint X-ray sources; even with observing times approaching $\sim 10^6$ s, there are insufficient (typically 10-100) counts to characterize the spectra of a source as that of a SNR or to measure the spatial extent.

Many SNRs have been identified as soft X-ray sources that are spatially coincident with objects in an independently derived catalog.  Alternatively, ``forced photometry'' may also be used to extract the X-ray fluxes (or upper limits) for SNRs that have been identified optically. For M33, Long10 used forced photometry on the {\em Chandra} ACIS data of the 137  then-known SNRs to report 2$\sigma$ detections of 82 objects, 58 of these at  $>3\sigma$.  XMM-{\em Newton} has greater sensitivity than {\em Chandra} and a larger field of view, but its   angular resolution is not high enough to resolve SNRs in confused regions.  \cite{garofali17} used forced photometry to detect 105 objects in the larger SNR sample discussed here at 3$\sigma$.  The typical X-ray luminosity of the objects detected by  \cite{garofali17} was \EXPU{7}{34}{\LUM} (0.2 - 2 keV) at the distance of M33.  In total, there are 112 objects which were detected at 3$\sigma$ in one or both X-ray studies.  We take these to be the X-ray detected sample of SNRs in M33, and have recorded this in Table \ref{table_snr_obs}.   

Of these objects, 106 also have \sii:\ha\ ratios of greater than 0.4.\footnote{The six outliers are L10-035, L10-040, LL14-005, LL14-008, LL14-134, and LL14-162. Of these, L10-035 and L10-040 had previously published spectra. For L10-035, Long10 reported a ratio of 0.36, not too different from what we report now, so this object does not appear to satisfy the formal criterion of 0.4 for a SNR. For L10-040 the spectrum is from \cite{smith93} who found a \sii:\ha\ ratio of 0.65, whereas we find 0.33.  L10-040 has a fairly complete shell 55 pc in diameter; it is probable that the difference in ratios is due to the placement of the Hectospec fiber compared to the slit used by \cite{smith93}.  At any rate new spectra of all of these objects would be desirable.}
It may be tempting to select these 106 objects as bona fide SNRs, and objects which have only a high \sii:\ha\ ratio as more suspect.  On whole, that is a view point which we share.  However, one should be aware that while these 106 objects may be the purest subsample of SNRs among the candidates, many Galactic SNRs would not have been detected in X-rays at the distance of M33, and so this subsample is by no means complete. Furthermore, given that many of the X-ray detections are near 3$\sigma$, and that at this limit it is sometimes difficult to separate a SNR from a peak in the X-ray background, X-ray detection is not always sufficient to authenticate an SNR candidate (see discussion in L10). 

Radio observations offer an additional means of verifying SNR candidates.  Unfortunately, the most recent radio survey of SNRs in M33 was by \citet{gordon99}, which targeted 98 SNR candidates that had been identified in \cite{gordon98}.  Of that list, \citet{gordon99} claimed radio detections of 53.  A new radio survey of M33 using JVLA observations is currently underway \cite[see][R. L. White et al.
 2018, in preparation]{long_crete} and should be a major step forward both in confirming candidate objects and in elucidating the relationships among X-ray, optical and radio emission from SNRs.

\subsection{Variation of Line Ratios with Galactocentric Distance}

\begin{figure}
\plottwo{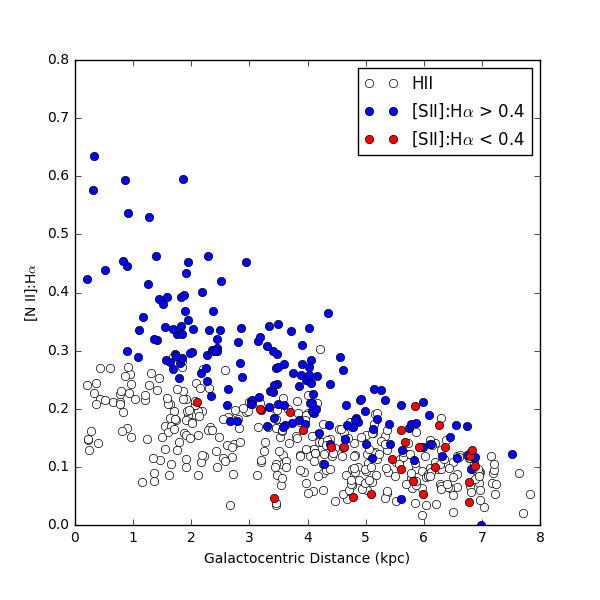}{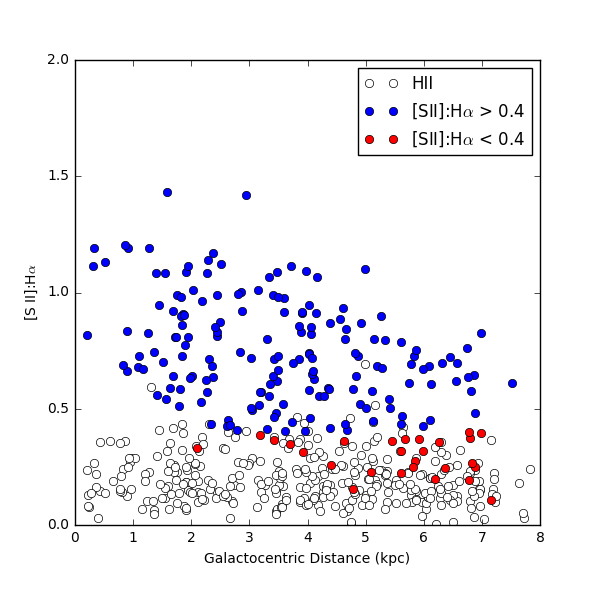}
\caption{The \nii:\ha\ ratios (left) and \sii:\ha\ ratios (right) as a function of galactocentric distance. In this case we have separated the SNR candidates into to those which have \sii:\ha\ ratios greater than or less than 0.4.  The \hii\ region ratios are also plotted. \label{fig_n2_galcen} }
\end{figure}

The large number and high quality of spectra now available for both SNRs and \hii\ regions in M33 allows us to revisit the situation with observed line ratios as a function of galactocentric distance (GCD)\@.
Fig.\ \ref{fig_n2_galcen} shows the variation of both \nii:\ha\ and \sii:\ha\ as a function of GCD, with different symbols for the SNR samples and \hii\ regions.\footnote{We note for the record that a similar plot for \oi:\ha\ shows no trend with GCD for either SNRs or \hii\ regions.} Three general aspects are immediately obvious: 1) the ratios are generally significantly higher in the SNR sample compared with \hii\ regions; 2) this separation is more dramatic near the inner part of the galaxy (i.e., there is a stronger gradient in line ratio with GCD for the SNRs, and more potential confusion at larger GCDs); and 3) there is a significant dispersion in ratio values within each object class at a given GCD. These trends have been noted previously, although the larger sample size here makes them more obvious.

The trends of the two line ratios with GCD show somewhat different behavior for SNRs compared with \hii\ regions and for the two line ratios themselves. For the SNR sample, both line ratios show a gradient, with decreasing values at larger GCD, despite the considerable dispersion.  For the \hii\ regions, a more modest gradient is seen in \nii:\ha, and it is not clear that any gradient is present for the \sii:\ha\ ratio.  The absence of a gradient in the \sii:\ha\ ratio in H II regions is probably due to the fact that most S in H II regions is more highly ionized; decreasing [S III] and [S IV] to hydrogen line ratios are observed in the IR and have been used to imply a S abundance gradient of d\,log(S/H)/d$R$ of -0.052$\pm$0.021 dex kpc$^{-1}$ \citep{rubin08}. The general offset in each ratio between the SNRs and \hii\ regions has been seen many times before and is attributable to the differing excitation mechanisms between the two classes of objects \citep[cf.][] {blair82,kewley02,allen08}. 

In addition to showing trends, it is quite apparent in the large samples of both \hii\ regions and  SNRs that there is very considerable dispersion in ratio values at a given GCD. Given the high quality of the spectra, most of this  dispersion must arise either from abundance variations or from varying physical conditions within the emitting plasmas of each type of object, or some combination the two. Since most, if not all of the SNRs identified in M33 have swept up much more mass from the ISM than was ejected by the SN, and since none of the spectra show direct evidence of ejecta, the actual abundances (and any variations) in the H II regions and SNRs should be similar, reflecting that of the ISM.

\citet{Rosolowsky08} and \citet{Magrini10} have both investigated \hii\ region abundances in M33 with significant (albeit somewhat smaller) samples of \hii\ regions (and planetary nebulae in the case of the latter paper), and found evidence for both overall abundance gradients and abundance variations (primarily O abundances, but also N) within averaged GCD bins. (Sulfur abundances were not addressed.) \cite{lin17} used their extensive observations to derive an inverse temperature gradient and the O abundance gradient in M33 using several different diagnostic ratios.  A similar analysis is also available for a large sample of \hii\ regions in M31 that also show variation in derived abundances at a given GCD  \citep{Sanders12}.  Both sets of authors claim a certain amount of variation in derived abundances at a given GCD, with \citet{Rosolowsky08} claiming azimuthal variation within a given GCD bin (but cf.\ \citet{Bresolin11} and \citet{Magrini10} showing that bright giant \hii\ regions have a steeper abundance gradient than the remainder of their sample). Unless one wants to believe that the actual abundances are different in these different objects, the inference here may be that ionization differences between the bright \hii\ region sample and fainter, more normal \hii\ regions are responsible for different derived abundance values.

The key issue in interpreting Fig.\ \ref{fig_n2_galcen} is the nature of the dispersion in ratios observed in the different classes of objects. Large grids of \hii\ region models, such as those of \cite{kewley02} and \cite{ValeAsari16} show that variations in ionization parameter can affect the observed line ratios, but  that abundance variations can impact observed ratios as well. Likewise, for shock models \citep{allen08}, we recently conducted a careful assessment for M83 SNRs \citep{winkler17} and concluded that both abundance variations and variations in reasonable assumptions for the shock conditions could contribute to the observed spread in line ratios.  

From the results of \citet{Magrini10}, it is tempting to assign a surface brightness variation as a surrogate for mean ionization of an \hii\ region.  Referring back to Fig.\ \ref{fig_lin} (right), the increase in \sii:\ha\ ratio at lower surface brightnesses may be a manifestation of the same effect, and is what is causing the uncertainty in object identifications based on this ratio.  However, as seen in Fig.~\ref{fig_lin} (left), the \nii:\ha\ ratio versus surface brightness shows some dispersion but a much smaller trend with surface brightness, indicating that little of the \nii:\ha\ dispersion can be attributed to this same effect.   

\begin{figure}
\plottwo{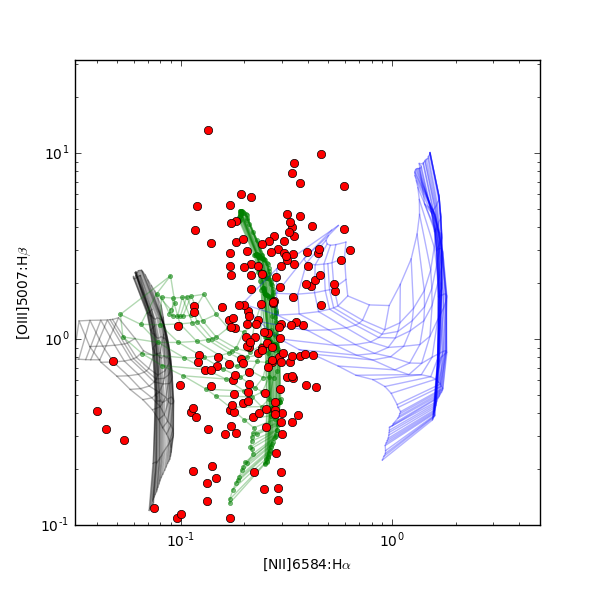}{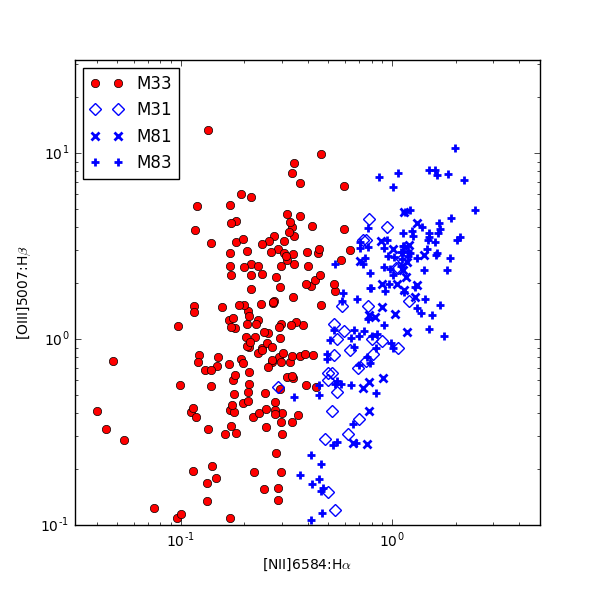}
\caption{Left:  Observed and model \oiii\,$\lambda$5007:\hb\ ratio as function of the \nii\,$\lambda$6583:\ha\ line ratio for SNRs and SNR candidates with spectra.  As discussed in the text, the black, green and blue meshes correspond to shock models with a range of shock velocities and pre-shock magnetic fields, and with metallicities corresponding to the SMC, LMC, and Milky Way, respectively.  Right: Observed ratios of the same lines for SNRs and SNR candidate samples in M31, M81, M83 as well as M33.  The ratios for M31, M81 and M83 were taken from \cite{galarza99}, \cite{lee15_m81} and \cite{winkler17}, respectively. \label{fig_n2o3} }
\end{figure}

In an attempt to shed some light on whether physical conditions or abundance variations explain the variations in the line ratios seen in the spectra of SNRs, we have compared the observed line ratios to the ratios shown in the grid of MAPPINGS III shock models presented by \cite{allen08}. An example of this comparison is shown in  the left panel of Fig.\ \ref{fig_n2o3}  for \oiii\,$\lambda$5007:\hb\ as a function of \nii\,$\lambda$6584:\ha.  For reasons discussed below, the grids we have elected to show are for shocks without precursors for models with Small Magellanic Cloud (SMC), Large Magellanic Cloud (LMC), and solar abundances.\footnote{In the nomenclature of \cite{allen08}, these are grids P, Q, and T for the SMC, LMC and Galaxy, respectively.}  Each of the grids is for a shock propagating into an ISM with density of 1 cm$^{-3}$.  Each grid covers a range of shock velocities from 100 - 1000 $\VEL$ and a range of pre-shock magnetic fields from \POW{-4} to 10 $\mu$G.  As such, the model grids were intended to cover the range of plausible conditions for radiative shocks propagating into an ISM with different metallicities.  There will, of course, be some old SNRs with shock speeds below 100 $\VEL$, and those may account for some of the fainter objects with weak [O III] emission.  Also, SNRs with primary shocks faster than 500 $\VEL$ are unlikely to have reached the radiative stage, so unless the primary shock has driven secondary shocks into denser knots of the ISM, they may be underrepresented in the sample. 


As shown in Fig.\ \ref{fig_n2o3}, there is a clear separation in the three model grids in the \nii:\ha\ ratios, while variation in terms of  \oiii:\hb\ is not as great. The line ratios that we observe from SNRs and SNR candidates in M33 show considerable scatter but cluster around the values seen in LMC grid.  The log of the O abundance relative to H for M33 is -3.6 to -3.7 \cite[e.g][]{lin17}; the O abundance in the LMC grid is -3.65, so it is comforting that the observations cluster near the LMC grid.  There are 3.8 times as many O atoms in the grid with solar abundances and 2.1 times fewer in the grid with SMC abundances.  The fact that there is an overall trend toward weaker \nii:\ha\ ratios with GC distance is consistent with there being an abundance gradient in M33.  

The left panel of Fig.\ \ref{fig_s2n2} shows a similar comparison for the ratio of \nii:\ha\ as a function of the ratio of \sii:\ha.  The same three model grids are 
shown.  Once again most of the data clusters near the LMC model grid.  If one concentrates on just the distribution along the \sii:\ha\ axis, it is fairly clear that much more of the variation in the expected \sii:\ha\ ratios arises from differences in the models; i.e., differences in physical conditions rather than in the abundance.  

As mentioned above, Figure~\ref{fig_n2o3} compares our line ratios with sets of \citet{allen08} models that do not include the contribution of the photoionization precursor.  This is because the models that include the calculated precursor give poor agreement with the observations, in the sense that the models including the precursor predict \oiii:\hb\ ratios higher than observed by a factor of 2 or 3.  This is because the low density in the precursor and the high ionizing flux from the shock imply a relatively high ionization state, and the emission is similar to that of an \hii\ region.  

The apparent weakness of the precursor contribution requires some explanation, as shocks faster than about 100 $\VEL$ produce substantial fluxes of ionizing photons.  The precursors produced by the ionizing flux are observed as faint, diffuse, emission outside several Galactic SNRs \citep[e.g.,][]{medina14}, but the surface brightness is so low that they make a miniscule contribution to the spectra from the bright, radiative SNR filaments.  SNRs in the LMC are observed at much lower effective spatial resolution, so the relative contribution of the precursor can be correspondingly larger. \citet{vancura92} separated the pre-shock and post-shock components of the lines of N49 based on the velocity shifts and widths, and found that 18\% of the \oiii\ and 5\% of \ha\ originate in the precursor.  At the still greater distance of M33, one might expect even more of the precursor emission would be captured in the 6 pc effective size of the Hectospec fibers.

The reason that the precursor contribution is weaker than the model prediction is that the models assume planar geometry and steady state emission.  Cox (1972) showed that a typical SNR begins to radiate a significant part of its energy when the shock speed slows to around 300 $\VEL$.  It then emits a huge burst of ionizing photons as the speed declines further to around 100 $\VEL$.  The cooling and recombination time in the photoionized gas is long compared to the time over which the SNR evolves, so only a fraction of the precursor emission is produced during the apparent lifetime of the SNR.  As the shock slows further still, shocks with speeds around 50 $\VEL$ moving in relic photoionization precursors should produce relatively faint emission with modest \oiii:\hb\ ratios, low densities, and small line widths.



\subsection{Comparison to SNR samples in Other Galaxies}

Although M33 may have the best studied SNR sample of any spiral galaxy, significant numbers of SNRs and SNR candidates have been identified in many other galaxies, including M31 
\cite[156,][]{lee14_m31} , M83 \cite[$\sim300$,][]{blair12,blair14}, and NGC~2403 \cite[149,][]{leonidaki13}, all using the \sii:\ha\ ratio as the primary criterion.  

\begin{figure}
\plottwo{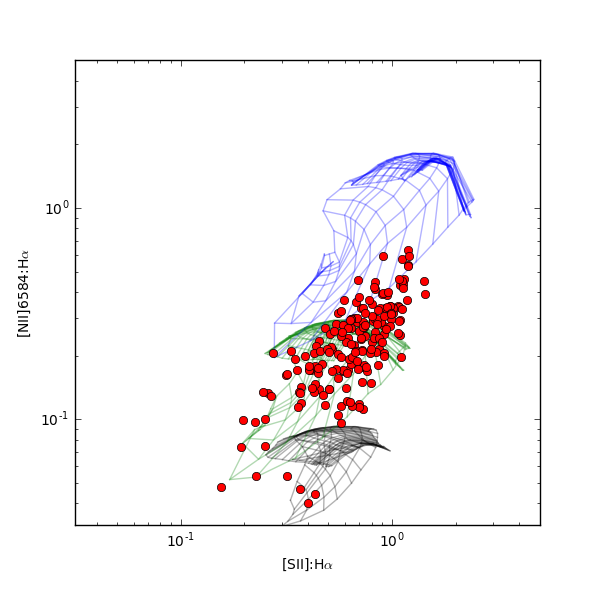}{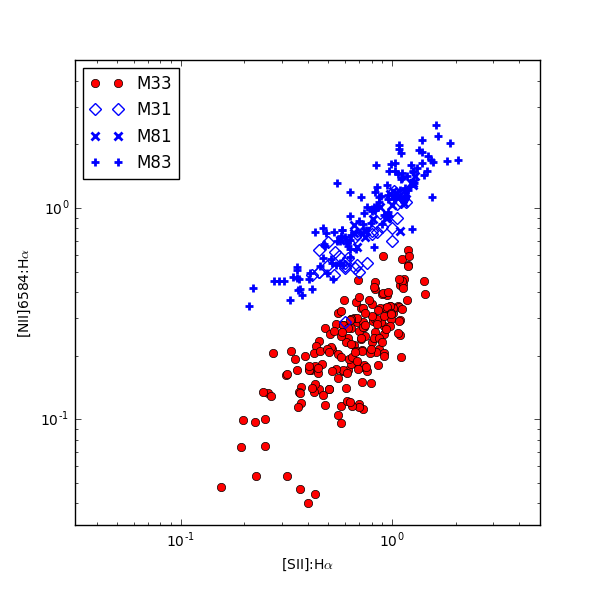}

\caption{Similar to Fig.\ \ref{fig_n2o3}, except here the \nii\,$\lambda$6583:\ha\ ratio is plotted as a function of the \sii:\ha\ ratio. \label{fig_s2n2} }
\end{figure}

For M83,  \cite{winkler17} obtained spectra of 118 candidates identified as having high \sii:\ha\ ratio from interference filter images and confirmed the high ratio in all but one.  With the notable exceptions of SN1957D \citep{long89, long12} and a very young SNR identified by \cite{blair15}, none of the optical spectra show broad lines expected from a Cas A analog.  This is true even though M83 has 41 objects (22 with spectra) with {\em HST}-measured diameters less than 0.5\arcsec (11 pc at $D = 4.6$ Mpc.  At a diameter of 11 pc, the primary shock velocity of a remnant from a \POW{51}{erg} explosion expanding into an ISM with density 1 cm$^{-3}$ would be $\sim 1800 \VEL$.  According to \cite{long14}, at least 87 of the approximately  300 optically identified SNR candidates in M83 have X-ray counterparts.  As in M33, smaller diameter candidates are more likely to be X-ray detected and to show higher densities from the ratios of the \sii\ lines.    As in M33, there is no evidence that the X-ray-detected objects differ from the non-detected objects in terms of reddening.  However, trends of \nii:\ha\ and \sii:\ha\  with galactocentric distance are less obvious in M83 than in M33, a fact that \cite{winkler17} argue is due to local abundance variations.  For a given \sii:\ha\ ratio, the \nii:\ha\ ratio is higher in M83 than M33, which almost certainly reflects the higher metallicity of M83.  This is illustrated in the right panels of Figures \ref{fig_n2o3} and \ref{fig_s2n2} where the data points taken from \cite{winkler17} are well separated from the locus of points for M33.

M83, where the original sample of SNR candidates was identified using interference filter images  taken in exquisite seeing ($\lesssim 0.5\arcsec$) from Magellan, and from {\em HST}, is an example of a galaxy where one can be reasonably certain that the majority of the objects in the sample are SNRs.   By contrast, NGC~2403, a spiral somewhat similar to M33 but at a distance of 3.2 Mpc, is an example of a galaxy where a large number of candidates have been identified, but where \citep [despite the best efforts of][] {leonidaki13} the sample is likely to be significantly contaminated by objects that are not actually SNRs.  To create their sample, \cite{leonidaki13} used images with seeing that ranged from 1.3\arcsec\ to 2.5\arcsec, corresponding to 20\,-\,40 pc at the distance of NGC2403, on a modest-sized (1.3 m) telescope.  They found 149 candidates, 102 of which had imaging-derived \sii:\ha\ ratios $> 0.4$, and 47 with ratios between 0.3 and 0.4.  They obtained spectra of 22 of the objects; while 7 of the 8 objects with imaging ratios  between 0.3 and 0.4 turned out to have spectroscopic \sii:\ha\ ratios greater than 0.4, only 5 of 14 objects with imaging ratios $>0.4$ had spectroscopic ratios $> 0.4$. 
Only about 40 ks of {\em Chandra} imaging exists for NGC~2403  (compared to 730 ks for M83), and so although \cite{leonidaki13} note that 6 of the 149 candidates are spatially coincident with X-ray sources in NGC~2403, all of these have hard X-ray spectra and are likely to be X-ray binaries (though \citealt{leonidaki13} also mention the possibility that these could be Crab-like SNRs).  Thus, whether or not a \sii:\ha\ ratio of 0.4 should qualify an emission nebula as a SNR candidate, near-complete samples of SNRs in nearby galaxies require high quality optical data with the best spatial resolution possible, spectroscopic follow-up on large telescopes, and if possible, deep X-ray observations at the resolution {\em Chandra} provides.

M31 is a galaxy where the quality of the existing SNR catalogs are probably in an intermediate state between the relative completeness of M33 and M83, and the incomplete state of  NGC~2403.  
Following early studies of M31 carried out with photographic plates, the first SNR search using CCDs, by \citet{braun93}, was limited to portions of the northwestern half of the galaxy.  A subsequent study by \citet{magnier95} covered a larger fraction of M31, but did not use continuum-subtracted images nor attempt any quantitative measurement of the \sii:\ha\ ratio.
By far the most thorough search for SNRs in M31 was done by \cite{lee14_m31}, in a companion study to the LL14 study of M33 that was similarly based on LGGS images.  They identified 156 SNR candidates, of which 76 were new to their study.  It may at first seem surprising that similar searches yielded over 25\% more candidates in M33 than in the much larger M31. 

As in M83 (but less so in M33) the SNR candidates in M31 are preferentially located in the spiral arms and ring structure, where most of the star formation is taking place.  Despite the fact that M31 is much more massive than M33, the star formation rates (SFR) in M31 \cite[0.4$\MSOL~\rm yr^{-1}$,][]{barmby06} and M33 \cite[0.45$\pm$0.1$\MSOL~\rm yr^{-1}$, ][]{verley09} are comparable.
One would expect the number of SNe to scale roughly as the SFR, and hence that M31 and M33 would have comparable numbers of SNRs.  The fact that surveys of M33 by Long10 and LL14, and of M31 by \citet{lee14_m31}, based on similar LGGS images of the two galaxies, have revealed $\sim 50\%$ more SNRs in M33 than in M31 is probably attributable to the fact that M31 is more inclined and more highly reddened than M33\@. \cite{lee14_m31} required all objects to have an integrated \sii:\ha\ ratio (from the LGGS images) strictly $> 0.4$ to be in their the catalog;  the mean ratio is 0.8.  As in the case of M83, the \sii:\ha\ ratios in M31 show only a slight gradient  with galactocentric distance, and a large scatter.  Only 23 of the objects in the M31 optical sample are in the XMM catalog of M31 SNRs presented by \cite{sasaki12}, but no one has yet carried out a study like that of  Long10 or \cite{garofali17} where the positions of the SNR candidates were searched for X-ray emission using forced photometry.  

The main question about the optical sample in M31 is similar to that for the M33 sample:  What fraction of the objects are actually SNRs?  The problem is worse in M31 than M33 though, because in the case of M31, there has been no systematic attempt to obtain spectra, and hence line ratios for the complete sample, and the surface brightness of the M31 sample \cite[see Fig. 18 of ][]{lee14_m31} extends to even fainter values than for M33.  Spectroscopic follow-up for the SNRs is needed to confirm the \sii:\ha\ ratios derived from imaging and to compare those ratios with \hii\ regions of similar surface brightness.  Spectra do exist for at least 33 SNRs in M31 \citep{galarza99}, and as in the case of M83, as shown in the right panel of Figures \ref{fig_n2o3} and \ref{fig_s2n2}, they lie in regions of the ratio diagrams expected for solar-like abundances.


\section{Summary \label{sec:summary}}

We have obtained spectra of 197 SNRs and SNR candidates in M33 using the Hectospec fiber-fed spectrograph at the 6.5m MMT Observatory at Mt.\ Hopkins.  These spectra cover the great majority of such objects listed in recent  catalogs by L10 and LL14.  Of these, 110 appear in both catalogs, while 18 are in L10 only and 69 are in LL14 only, for a total of 217.  These data were analyzed and compared with a subset of \hii\ regions recently published in Lin17.  Our principal results are as follows:

\begin{itemize}
\item
Fits to the emission-line spectra show that the flux ratio of \SiiL:\ha\ is above 0.4 for 170 of the 197 objects.  Traditionally, a value  for the \sii:\ha\ ratio $>0.4$ has been taken as the principal optical diagnostic for shock-heated material, such as that found in SNRs, while photoionized nebulae have usually been found to have significantly lower values.  If the 8 objects that we did not observe but for which archival spectra exist are taken into account, then 178 of the 217 objects proposed to be SNRs have \sii:\ha\ ratio $>0.4$.   The 39 objects which either have not been observed spectroscopically or which appear to have ratios less than 0.4 should be regarded as questionable, at least until adequate spectra are obtained.
\item
While the majority of the emission nebulae in the SNR candidate lists that have high \sii:\ha\ ratios are almost certainly SNRs, a comparison of the line ratios from the SNR sample to the \hii\ region sample of Lin17 shows  \hii\ region \sii:\ha\ ratios that are rising as the surface brightness decreases. Thus,  it is apparent that  the \sii:\ha\ ratio criterion alone is not completely reliable for distinguishing between these two classes of nebulae. The \oi:\ha\ line ratio provides further confirmation that many candidates with \sii:\ha\ $>0.4$ are indeed SNRs, but adds no new objects, primarily because the \oi\ lines are weaker than the \sii\ ones, and are confused with night sky emission. Furthermore, some of the fainter \hii\ regions have \oi:\ha\ ratios that overlap with those seen in comparably faint SNR candidates.
\item
In an attempt to develop a clean sample of SNRs from optical criteria only, another potential discriminant is the velocity broadening of the emission lines.  The bright \hii\ regions that we observed all have line widths that are consistent with the instrumental width, whereas widths we measure for the SNR candidates are measurably broader.  However, the spectral resolution of Hectospec ($\sim 5$ \AA) is not sufficient to effectively separate SNRs with shocks $\lesssim 200 \VEL$ from \hii\ regions.  Higher resolution spectra would, in principle, provide an effective discriminant, and would provide important information about shock velocities and the evolutionary state of the small diameter SNRs, but obtaining them with high S/N for faint objects would require a significant investment in telescope time.
\item
Of the 217 optical SNRs and candidates in the catalog compiled here, 112 have been detected at $3\sigma$ or higher in one or both of the recent deep  X-ray surveys from {\em Chandra} \citep{long10} and XMM-{\em Newton} \citep{garofali17}.  Of these 112 objects, 106 have \sii:\ha\ ratios greater than 0.4.
\item
The SNRs show a strong radial gradient in both the \nii:\ha\ and \sii:\ha\ ratios, decreasing at larger distances from the galaxy's center.  As previous studies that have observed similar gradients in M33 with far smaller samples have concluded, these are almost certainly due to decreasing elemental abundances with larger GCD.  The \hii\ region sample shows a similar but milder gradient  in \nii:\ha, but the \sii:\ha\ ratio is essentially flat with GCD.  
\item
The substantial samples of both SNRs and \hii\ regions with high quality spectra also allow us to confirm a large dispersion in these line ratios at a given galactocentric distance.  Comparison to models indicates that both varying physical conditions and varying abundances contribute to this spread in observed ratios.
\item
The line ratios seen in the SNR sample of M33 are consistent with those predicted by shock models expanding into an ISM with the metallicity of M33, but only if precursor radiation is modest.  Spectra of SNRs in other galaxies such as M31 and M83 have higher line ratios reflecting their higher metallicity.
\end{itemize}

\vspace{0.2 in}
In order to fully understand the  population of SNRs in M33 and other nearby galaxies, we would ideally like to compare SNR samples identified independently from deep surveys in the optical, X-ray, and radio bands. The current study is at least close to what can be achieved optically with current instruments.  One might try to create a ``gold sample" of optically identified SNRs, but this has the disadvantage that it would exclude a number of actual SNRs (especially fainter ones), so we have elected not to attempt this.
Deep observations from {\em Chandra} \citep{long10} and XMM-{\em Newton} \citep{garofali17} have successfully detected a majority of M33's SNRs,  though there are doubtless others whose X-ray flux falls below the detection threshold of those studies.  Deeper X-ray observations of M33 are unlikely to be forthcoming with the current generation of X-ray observatories, especially since the soft X-ray sensitivity of the ACIS instrument on {\em Chandra} has declined due to the build-up of contamination on the detector filter.  However, deeper and more complete radio studies than that of \cite{gordon99} are definitely practicable with current instruments, and we are currently pursuing this goal. 

\acknowledgments

Observations reported here were obtained at the MMT Observatory, a joint facility of the Smithsonian Institution and the University of Arizona.  PFW acknowledges support from the NSF through grant AST-1714281.  WPB acknowledges ongoing support from the JHU Center for Astrophysical Sciences. We also appreciate the support with Hectospec observations provided by Nelson Caldwell, and valuable comments made by Paul P.\ Plucinsky on the paper prior to submission.

\facilities{MMT (Hectospec)}
\software{astropy \citep{astropy}
}



\clearpage

\startlongtable
\begin{deluxetable}
{rccrrcccclcc}

\tablecaption{SNR Candidates in M33 
\label{table_snr_obs}
}
\tablehead{\colhead{Source} & 
 \colhead{RA} & 
 \colhead{Dec} & 
 \colhead{Diameter} & 
 \colhead{$\rho$} &
 \colhead{L10} & 
 \colhead{LL14} & 
 \colhead{XMM} & 
 \colhead{X-ray} & 
 \colhead{Previous} & 
 \colhead{New} & 
 \colhead{[S\,II]:H$\alpha$} 
\\
\colhead{~} & 
 \colhead{(2000)} & 
 \colhead{(2000)} & 
 \colhead{(pc)} & 
 \colhead{(kpc)} & 
 \colhead{~} & 
 \colhead{~} & 
 \colhead{-} & 
 \colhead{Detected} & 
 \colhead{Spectrum} & 
 \colhead{Spectrum} & 
 \colhead{Confirmed} 
}
\tabletypesize{\tiny}
\tablewidth{0pt}\startdata
L10-001 &  01:32:30.37 &  30:27:46.9 &  126 &  6.5 &  L10-001 &  -- &  XMM-003 &  yes &  G98 &  yes &  yes \\ 
L10-002 &  01:32:31.41 &  30:35:32.9 &  33 &  6.5 &  L10-002 &  LL14-003 &  XMM-004 &  yes &  G98 &  yes &  yes \\ 
L10-003 &  01:32:42.54 &  30:20:58.9 &  104 &  6.1 &  L10-003 &  -- &  XMM-011 &  no &  G98 &  yes &  yes \\ 
L10-004 &  01:32:44.83 &  30:22:14.6 &  42 &  5.8 &  L10-004 &  LL14-011 &  XMM-013 &  no &  -- &  yes &  yes \\ 
L10-005 &  01:32:46.73 &  30:34:37.8 &  49 &  5.2 &  L10-005 &  LL14-013 &  XMM-015 &  yes &  L10 &  yes &  yes \\ 
L10-006 &  01:32:52.76 &  30:38:12.6 &  60 &  4.9 &  L10-006 &  LL14-015 &  XMM-017 &  yes &  S93 &  yes &  yes \\ 
L10-007 &  01:32:53.36 &  30:48:23.1 &  77 &  6.3 &  L10-007 &  LL14-018 &  XMM-019 &  yes &  -- &  yes &  yes \\ 
L10-008 &  01:32:53.40 &  30:37:56.9 &  55 &  4.8 &  L10-008 &  LL14-017 &  XMM-020 &  no &  S93 &  yes &  yes \\ 
L10-009 &  01:32:54.10 &  30:25:31.8 &  42 &  4.9 &  L10-009 &  LL14-019 &  XMM-021 &  no &  -- &  yes &  yes \\ 
L10-010 &  01:32:56.15 &  30:40:36.4 &  97 &  4.9 &  L10-010 &  LL14-021 &  XMM-022 &  yes &  S93 &  yes &  yes \\ 
L10-011 &  01:32:57.07 &  30:39:27.1 &  23 &  4.7 &  L10-011 &  LL14-022 &  XMM-024 &  yes &  L10 &  yes &  yes \\ 
L10-012 &  01:33:00.15 &  30:30:46.2 &  56 &  4.1 &  L10-012 &  -- &  XMM-026 &  yes &  -- &  yes &  yes \\ 
L10-013 &  01:33:00.42 &  30:44:08.1 &  37 &  5.0 &  L10-013 &  LL14-024 &  XMM-027 &  yes &  G98 &  yes &  yes \\ 
L10-014 &  01:33:00.67 &  30:30:59.3 &  49 &  4.1 &  L10-014 &  LL14-025 &  XMM-028 &  no &  -- &  yes &  yes \\ 
L10-015 &  01:33:01.51 &  30:30:49.6 &  32 &  4.0 &  L10-015 &  LL14-026 &  XMM-029 &  no &  -- &  no &  - \\ 
L10-016 &  01:33:02.93 &  30:32:29.6 &  55 &  3.9 &  L10-016 &  LL14-027 &  XMM-030 &  yes &  L10 &  yes &  yes \\ 
L10-017 &  01:33:03.57 &  30:31:20.9 &  36 &  3.8 &  L10-017 &  LL14-028 &  XMM-031 &  yes &  G98 &  no &  yes \\ 
L10-018 &  01:33:04.03 &  30:39:53.7 &  34 &  4.1 &  L10-018 &  LL14-029 &  XMM-032 &  yes &  S93 &  yes &  yes \\ 
L10-019 &  01:33:07.55 &  30:42:52.5 &  74 &  4.2 &  L10-019 &  -- &  XMM-033 &  yes &  -- &  yes &  yes \\ 
L10-020 &  01:33:08.98 &  30:26:58.9 &  55 &  3.9 &  L10-020 &  LL14-031 &  XMM-035 &  yes &  -- &  yes &  yes \\ 
L10-021 &  01:33:09.87 &  30:39:34.9 &  70 &  3.6 &  L10-021 &  LL14-033 &  XMM-037 &  no &  S93 &  no &  yes \\ 
L10-022 &  01:33:10.18 &  30:42:22.0 &  31 &  3.9 &  L10-022 &  LL14-034 &  XMM-038 &  yes &  L10 &  yes &  yes \\ 
L10-023 &  01:33:11.10 &  30:39:43.7 &  29 &  3.5 &  L10-023 &  LL14-035 &  XMM-039 &  yes &  L10 &  yes &  yes \\ 
L10-024 &  01:33:11.28 &  30:34:23.5 &  102 &  3.2 &  L10-024 &  LL14-036 &  XMM-040 &  no &  S93 &  yes &  yes \\ 
L10-025 &  01:33:11.76 &  30:38:41.5 &  28 &  3.3 &  L10-025 &  LL14-037 &  XMM-041 &  yes &  L10 &  yes &  yes \\ 
L10-026 &  01:33:16.73 &  30:46:10.3 &  77 &  4.0 &  L10-026 &  LL14-041 &  XMM-045 &  no &  -- &  yes &  yes \\ 
L10-027 &  01:33:17.44 &  30:31:28.5 &  48 &  2.9 &  L10-027 &  LL14-042 &  XMM-046 &  yes &  G98 &  yes &  yes \\ 
L10-028 &  01:33:18.80 &  30:27:04.4 &  183 &  3.5 &  L10-028 &  -- &  XMM-049 &  no &  -- &  yes &  yes \\ 
L10-029 &  01:33:18.94 &  30:46:51.9 &  69 &  4.0 &  L10-029 &  LL14-045 &  XMM-050 &  yes &  -- &  yes &  yes \\ 
L10-030 &  01:33:21.64 &  30:31:31.1 &  79 &  2.6 &  L10-030 &  LL14-050 &  XMM-055 &  no &  -- &  yes &  yes \\ 
L10-031 &  01:33:22.67 &  30:27:04.0 &  23 &  3.3 &  L10-031 &  LL14-052 &  XMM-057 &  yes &  L10 &  yes &  yes \\ 
L10-032 &  01:33:23.85 &  30:26:13.5 &  25 &  3.5 &  L10-032 &  LL14-053 &  XMM-058 &  yes &  BK85 &  yes &  yes \\ 
L10-033 &  01:33:27.07 &  30:47:48.6 &  70 &  3.6 &  L10-033 &  LL14-056 &  XMM-061 &  no &  L10 &  yes &  yes \\ 
L10-034 &  01:33:28.08 &  30:31:35.0 &  36 &  2.3 &  L10-034 &  LL14-060 &  XMM-065 &  yes &  G98 &  yes &  yes \\ 
L10-035 &  01:33:28.96 &  30:47:43.5 &  22 &  3.4 &  L10-035 &  -- &  XMM-066 &  yes &  L10 &  yes &  no \\ 
L10-036 &  01:33:29.05 &  30:42:17.0 &  22 &  2.3 &  L10-036 &  LL14-061 &  XMM-067 &  yes &  S93 &  yes &  yes \\ 
L10-037 &  01:33:29.45 &  30:49:10.8 &  35 &  3.8 &  L10-037 &  LL14-062 &  XMM-068 &  yes &  G98 &  yes &  yes \\ 
L10-038 &  01:33:30.21 &  30:47:43.8 &  55 &  3.4 &  L10-038 &  LL14-064 &  XMM-070 &  no &  L10 &  yes &  yes \\ 
L10-039 &  01:33:31.25 &  30:33:33.4 &  16 &  1.9 &  L10-039 &  LL14-067 &  XMM-073 &  yes &  L10 &  yes &  yes \\ 
L10-040 &  01:33:31.34 &  30:42:18.3 &  58 &  2.1 &  L10-040 &  LL14-068 &  XMM-074 &  yes &  S93 &  yes &  no \\ 
L10-041 &  01:33:31.80 &  30:31:01.1 &  100 &  2.3 &  L10-041 &  LL14-069 &  XMM-075 &  yes &  -- &  yes &  yes \\ 
L10-042 &  01:33:35.14 &  30:23:07.5 &  46 &  4.1 &  L10-042 &  LL14-071 &  XMM-078 &  yes &  -- &  yes &  yes \\ 
L10-043 &  01:33:35.39 &  30:42:32.1 &  88 &  1.8 &  L10-043 &  -- &  XMM-079 &  yes &  -- &  yes &  yes \\ 
L10-044 &  01:33:35.61 &  30:49:23.0 &  32 &  3.4 &  L10-044 &  LL14-073 &  XMM-080 &  yes &  L10 &  yes &  yes \\ 
L10-045 &  01:33:35.90 &  30:36:27.4 &  33 &  1.2 &  L10-045 &  LL14-074 &  XMM-082 &  yes &  L10 &  yes &  yes \\ 
L10-046 &  01:33:37.09 &  30:32:53.5 &  42 &  1.8 &  L10-046 &  LL14-076 &  XMM-083 &  yes &  L10 &  yes &  yes \\ 
L10-047 &  01:33:37.75 &  30:40:09.2 &  53 &  1.2 &  L10-047 &  LL14-077 &  XMM-085 &  yes &  S93 &  no &  yes \\ 
L10-048 &  01:33:38.01 &  30:42:18.2 &  25 &  1.6 &  L10-048 &  LL14-078 &  XMM-086 &  no &  S93 &  yes &  yes \\ 
L10-049 &  01:33:40.66 &  30:39:40.8 &  46 &  0.9 &  L10-049 &  LL14-082 &  XMM-091 &  yes &  -- &  yes &  yes \\ 
L10-050 &  01:33:40.73 &  30:42:35.7 &  75 &  1.4 &  L10-050 &  -- &  XMM-092 &  no &  S93 &  yes &  yes \\ 
L10-051 &  01:33:40.87 &  30:52:13.7 &  64 &  3.9 &  L10-051 &  LL14-083 &  XMM-093 &  no &  G98 &  yes &  yes \\ 
L10-052 &  01:33:41.30 &  30:32:28.4 &  37 &  1.8 &  L10-052 &  LL14-084 &  XMM-094 &  no &  -- &  yes &  yes \\ 
L10-053 &  01:33:41.71 &  30:21:04.1 &  38 &  4.8 &  L10-053 &  LL14-085 &  XMM-096 &  no &  G98 &  yes &  yes \\ 
L10-054 &  01:33:42.24 &  30:20:57.8 &  49 &  4.8 &  L10-054 &  LL14-086 &  XMM-097 &  no &  -- &  yes &  yes \\ 
L10-055 &  01:33:42.91 &  30:41:49.5 &  48 &  1.1 &  L10-055 &  LL14-087 &  XMM-098 &  yes &  -- &  yes &  yes \\ 
L10-056 &  01:33:43.49 &  30:41:03.8 &  27 &  0.9 &  L10-056 &  LL14-088 &  XMM-099 &  yes &  G98 &  no &  yes \\ 
L10-057 &  01:33:43.70 &  30:36:11.5 &  40 &  0.9 &  L10-057 &  LL14-089 &  XMM-100 &  yes &  S93 &  yes &  yes \\ 
L10-058 &  01:33:45.26 &  30:32:20.1 &  71 &  1.8 &  L10-058 &  LL14-090 &  XMM-101 &  no &  -- &  yes &  yes \\ 
L10-059 &  01:33:47.46 &  30:39:44.7 &  45 &  0.3 &  L10-059 &  LL14-091 &  XMM-102 &  yes &  -- &  yes &  yes \\ 
L10-060 &  01:33:48.35 &  30:39:28.4 &  17 &  0.2 &  L10-060 &  LL14-095 &  XMM-106 &  no &  G98 &  yes &  yes \\ 
L10-061 &  01:33:48.50 &  30:33:07.9 &  63 &  1.7 &  L10-061 &  -- &  XMM-107 &  yes &  S93 &  yes &  yes \\ 
L10-062 &  01:33:49.75 &  30:30:49.7 &  76 &  2.4 &  L10-062 &  LL14-097 &  XMM-108 &  no &  -- &  yes &  yes \\ 
L10-063 &  01:33:49.90 &  30:30:16.7 &  57 &  2.5 &  L10-063 &  LL14-098 &  XMM-109 &  no &  -- &  yes &  yes \\ 
L10-064 &  01:33:50.12 &  30:35:28.6 &  52 &  1.1 &  L10-064 &  LL14-099 &  XMM-110 &  yes &  S93 &  yes &  yes \\ 
L10-065 &  01:33:51.06 &  30:43:56.2 &  53 &  1.2 &  L10-065 &  LL14-100 &  XMM-111 &  yes &  S93 &  yes &  yes \\ 
L10-066 &  01:33:51.67 &  30:30:59.6 &  63 &  2.4 &  L10-066 &  LL14-101 &  XMM-112 &  yes &  G98 &  yes &  yes \\ 
L10-067 &  01:33:51.71 &  30:30:43.4 &  49 &  2.5 &  L10-067 &  LL14-102 &  XMM-113 &  no &  -- &  yes &  yes \\ 
L10-068 &  01:33:52.15 &  30:56:33.4 &  112 &  4.6 &  L10-068 &  -- &  XMM-114 &  yes &  -- &  yes &  yes \\ 
L10-069 &  01:33:54.28 &  30:33:47.9 &  51 &  1.7 &  L10-069 &  LL14-104 &  XMM-116 &  yes &  S93 &  yes &  yes \\ 
L10-070 &  01:33:54.51 &  30:45:18.7 &  24 &  1.4 &  L10-070 &  LL14-105 &  XMM-118 &  yes &  G98 &  yes &  yes \\ 
L10-071 &  01:33:54.91 &  30:33:11.0 &  24 &  1.9 &  L10-071 &  LL14-107 &  XMM-119 &  yes &  S93 &  yes &  yes \\ 
L10-072 &  01:33:55.01 &  30:39:57.3 &  36 &  0.3 &  L10-072 &  LL14-108 &  XMM-120 &  no &  -- &  yes &  yes \\ 
L10-073 &  01:33:56.49 &  30:21:27.0 &  61 &  5.2 &  L10-073 &  LL14-110 &  XMM-122 &  yes &  -- &  yes &  yes \\ 
L10-074 &  01:33:56.97 &  30:34:58.7 &  34 &  1.6 &  L10-074 &  LL14-111 &  XMM-123 &  yes &  L10 &  yes &  yes \\ 
L10-075 &  01:33:57.13 &  30:40:48.5 &  48 &  0.5 &  L10-075 &  LL14-112 &  XMM-124 &  no &  G98 &  yes &  yes \\ 
L10-076 &  01:33:57.13 &  30:35:06.1 &  24 &  1.5 &  L10-076 &  LL14-113 &  XMM-125 &  no &  -- &  yes &  yes \\ 
L10-077 &  01:33:58.06 &  30:32:09.6 &  27 &  2.3 &  L10-077 &  LL14-115 &  XMM-127 &  yes &  S93 &  yes &  yes \\ 
L10-078 &  01:33:58.07 &  30:37:54.6 &  20 &  0.9 &  L10-078 &  LL14-116 &  XMM-128 &  yes &  L10 &  yes &  yes \\ 
L10-079 &  01:33:58.15 &  30:48:36.4 &  62 &  2.3 &  L10-079 &  -- &  XMM-130 &  yes &  -- &  yes &  yes \\ 
L10-080 &  01:33:58.42 &  30:36:24.3 &  11 &  1.3 &  L10-080 &  LL14-117 &  XMM-129 &  yes &  L10 &  yes &  yes \\ 
L10-081 &  01:33:58.51 &  30:33:32.3 &  38 &  2.0 &  L10-081 &  LL14-119 &  XMM-132 &  yes &  S93 &  yes &  yes \\ 
L10-082 &  01:33:58.52 &  30:51:54.3 &  60 &  3.1 &  L10-082 &  LL14-118 &  XMM-131 &  yes &  G98 &  yes &  yes \\ 
L10-083 &  01:33:59.93 &  30:34:21.2 &  34 &  1.9 &  L10-083 &  LL14-121 &  XMM-134 &  no &  G98 &  no &  yes \\ 
L10-084 &  01:34:00.31 &  30:42:19.4 &  36 &  0.9 &  L10-084 &  LL14-123 &  XMM-136 &  yes &  L10 &  yes &  yes \\ 
L10-085 &  01:34:00.32 &  30:47:24.1 &  46 &  1.9 &  L10-085 &  LL14-124 &  XMM-135 &  yes &  -- &  yes &  yes \\ 
L10-086 &  01:34:00.60 &  30:49:04.2 &  17 &  2.4 &  L10-086 &  LL14-126 &  XMM-139 &  yes &  L10 &  yes &  yes \\ 
L10-087 &  01:34:01.34 &  30:35:20.2 &  51 &  1.7 &  L10-087 &  LL14-127 &  XMM-140 &  yes &  L10 &  yes &  yes \\ 
L10-088 &  01:34:02.24 &  30:31:06.8 &  64 &  2.9 &  L10-088 &  LL14-129 &  XMM-142 &  yes &  S93 &  yes &  yes \\ 
L10-089 &  01:34:03.31 &  30:36:22.9 &  96 &  1.6 &  L10-089 &  LL14-130 &  XMM-143 &  yes &  -- &  yes &  yes \\ 
L10-090 &  01:34:03.48 &  30:44:43.8 &  45 &  1.4 &  L10-090 &  LL14-131 &  XMM-144 &  yes &  S93 &  yes &  yes \\ 
L10-091 &  01:34:04.26 &  30:32:57.1 &  40 &  2.5 &  L10-091 &  LL14-132 &  XMM-145 &  yes &  -- &  yes &  yes \\ 
L10-092 &  01:34:07.23 &  30:36:22.0 &  105 &  1.9 &  L10-092 &  LL14-135 &  XMM-148 &  yes &  S93 &  yes &  yes \\ 
L10-093 &  01:34:07.50 &  30:37:08.0 &  23 &  1.8 &  L10-093 &  LL14-136 &  XMM-149 &  yes &  L10 &  yes &  yes \\ 
L10-094 &  01:34:08.37 &  30:46:33.2 &  23 &  1.9 &  L10-094 &  LL14-138 &  XMM-151 &  yes &  L10 &  yes &  yes \\ 
L10-095 &  01:34:10.02 &  30:47:14.9 &  26 &  2.1 &  L10-095 &  LL14-139 &  XMM-152 &  yes &  L10 &  no &  yes \\ 
L10-096 &  01:34:10.70 &  30:42:24.0 &  22 &  1.6 &  L10-096 &  LL14-140 &  XMM-153 &  yes &  S93 &  yes &  yes \\ 
L10-097 &  01:34:11.04 &  30:38:59.9 &  18 &  1.8 &  L10-097 &  LL14-141 &  XMM-154 &  yes &  G98 &  yes &  yes \\ 
L10-098 &  01:34:12.69 &  30:35:12.0 &  70 &  2.6 &  L10-098 &  -- &  XMM-157 &  no &  S93 &  yes &  yes \\ 
L10-099 &  01:34:13.02 &  30:48:36.1 &  55 &  2.4 &  L10-099 &  LL14-145 &  XMM-159 &  yes &  -- &  yes &  yes \\ 
L10-100 &  01:34:13.65 &  30:43:27.0 &  29 &  1.8 &  L10-100 &  LL14-146 &  XMM-160 &  yes &  S93 &  yes &  yes \\ 
L10-101 &  01:34:13.71 &  30:48:17.5 &  61 &  2.4 &  L10-101 &  LL14-147 &  XMM-161 &  yes &  -- &  yes &  yes \\ 
L10-102 &  01:34:14.10 &  30:34:30.9 &  42 &  2.9 &  L10-102 &  LL14-149 &  XMM-163 &  yes &  S93 &  no &  yes \\ 
L10-103 &  01:34:14.35 &  30:41:53.6 &  51 &  1.9 &  L10-103 &  LL14-150 &  XMM-164 &  no &  S93 &  yes &  yes \\ 
L10-104 &  01:34:14.38 &  30:39:41.6 &  42 &  2.0 &  L10-104 &  LL14-151 &  XMM-166 &  yes &  S93 &  yes &  yes \\ 
L10-105 &  01:34:14.41 &  30:53:51.9 &  53 &  3.6 &  L10-105 &  LL14-152 &  XMM-165 &  yes &  G98 &  yes &  yes \\ 
L10-106 &  01:34:14.67 &  30:31:50.9 &  70 &  3.5 &  L10-106 &  LL14-154 &  XMM-168 &  yes &  -- &  yes &  yes \\ 
L10-107 &  01:34:15.57 &  30:32:59.9 &  37 &  3.3 &  L10-107 &  LL14-155 &  XMM-169 &  yes &  S93 &  yes &  yes \\ 
L10-108 &  01:34:16.31 &  30:52:32.7 &  80 &  3.3 &  L10-108 &  LL14-156 &  XMM-170 &  yes &  L10 &  yes &  yes \\ 
L10-109 &  01:34:17.00 &  30:51:47.1 &  29 &  3.2 &  L10-109 &  LL14-157 &  XMM-172 &  no &  L10 &  yes &  no \\ 
L10-110 &  01:34:17.03 &  30:33:57.7 &  57 &  3.2 &  L10-110 &  LL14-158 &  XMM-171 &  yes &  -- &  yes &  yes \\ 
L10-111 &  01:34:17.61 &  30:41:23.3 &  55 &  2.2 &  L10-111 &  LL14-159 &  XMM-173 &  yes &  S93 &  yes &  yes \\ 
L10-112 &  01:34:18.32 &  30:54:05.8 &  87 &  3.7 &  L10-112 &  LL14-160 &  XMM-174 &  no &  -- &  yes &  yes \\ 
L10-113 &  01:34:19.28 &  30:33:45.9 &  42 &  3.4 &  L10-113 &  LL14-161 &  XMM-175 &  no &  S93 &  yes &  yes \\ 
L10-114 &  01:34:19.87 &  30:33:56.0 &  26 &  3.4 &  L10-114 &  LL14-164 &  XMM-178 &  yes &  S93 &  yes &  yes \\ 
L10-115 &  01:34:23.23 &  30:25:24.9 &  51 &  5.6 &  L10-115 &  LL14-165 &  XMM-180 &  yes &  -- &  yes &  yes \\ 
L10-116 &  01:34:23.27 &  30:54:23.9 &  46 &  3.9 &  L10-116 &  LL14-166 &  XMM-179 &  yes &  L10 &  yes &  yes \\ 
L10-117 &  01:34:25.09 &  30:54:58.1 &  70 &  4.1 &  L10-117 &  LL14-169 &  XMM-183 &  yes &  L10 &  yes &  yes \\ 
L10-118 &  01:34:25.41 &  30:48:30.9 &  56 &  3.0 &  L10-118 &  LL14-170 &  XMM-184 &  yes &  G98 &  yes &  yes \\ 
L10-119 &  01:34:25.87 &  30:33:16.8 &  37 &  4.0 &  L10-119 &  LL14-171 &  XMM-185 &  yes &  -- &  yes &  yes \\ 
L10-120 &  01:34:29.61 &  30:41:33.4 &  47 &  3.2 &  L10-120 &  LL14-172 &  XMM-186 &  no &  G98 &  yes &  yes \\ 
L10-121 &  01:34:30.29 &  30:35:44.8 &  57 &  4.0 &  L10-121 &  LL14-175 &  XMM-188 &  yes &  G98 &  yes &  yes \\ 
L10-122 &  01:34:31.85 &  30:56:41.5 &  115 &  4.6 &  L10-122 &  -- &  XMM-189 &  yes &  -- &  yes &  yes \\ 
L10-123 &  01:34:32.41 &  30:35:32.6 &  44 &  4.2 &  L10-123 &  LL14-176 &  XMM-190 &  yes &  -- &  yes &  yes \\ 
L10-124 &  01:34:33.02 &  30:46:39.2 &  14 &  3.4 &  L10-124 &  LL14-177 &  XMM-191 &  yes &  BK85 &  yes &  yes \\ 
L10-125 &  01:34:35.41 &  30:52:12.7 &  42 &  4.0 &  L10-125 &  LL14-178 &  XMM-192 &  yes &  -- &  yes &  yes \\ 
L10-126 &  01:34:36.22 &  30:36:23.6 &  50 &  4.4 &  L10-126 &  LL14-179 &  XMM-193 &  no &  -- &  yes &  yes \\ 
L10-127 &  01:34:39.00 &  30:37:59.8 &  89 &  4.4 &  L10-127 &  LL14-181 &  XMM-195 &  yes &  -- &  yes &  yes \\ 
L10-128 &  01:34:40.73 &  30:43:36.4 &  26 &  4.0 &  L10-128 &  LL14-184 &  XMM-198 &  yes &  L10 &  yes &  yes \\ 
L10-129 &  01:34:41.10 &  30:43:28.3 &  43 &  4.1 &  L10-129 &  LL14-185 &  XMM-199 &  yes &  BK85 &  no &  yes \\ 
L10-130 &  01:34:41.23 &  30:43:55.4 &  38 &  4.1 &  L10-130 &  LL14-186 &  XMM-200 &  no &  -- &  yes &  yes \\ 
L10-131 &  01:34:41.89 &  30:37:35.2 &  159 &  4.7 &  L10-131 &  -- &  XMM-201 &  no &  -- &  yes &  yes \\ 
L10-132 &  01:34:44.62 &  30:42:38.8 &  58 &  4.4 &  L10-132 &  -- &  XMM-204 &  no &  -- &  yes &  no \\ 
L10-133 &  01:34:54.88 &  30:41:17.0 &  79 &  5.4 &  L10-133 &  -- &  XMM-210 &  no &  -- &  yes &  yes \\ 
L10-134 &  01:34:56.44 &  30:36:23.2 &  62 &  6.1 &  L10-134 &  LL14-194 &  XMM-211 &  no &  -- &  yes &  yes \\ 
L10-135 &  01:35:00.36 &  30:40:05.0 &  69 &  6.0 &  L10-135 &  LL14-197 &  XMM-214 &  no &  -- &  yes &  yes \\ 
L10-136 &  01:35:01.22 &  30:38:17.1 &  132 &  6.3 &  L10-136 &  -- &  XMM-216 &  no &  -- &  yes &  no \\ 
L10-137 &  01:35:03.18 &  30:37:09.6 &  130 &  6.6 &  L10-137 &  -- &  XMM-218 &  no &  -- &  yes &  yes \\ 
LL14-001 &  01:32:25.78 &  30:30:04.0 &  82 &  6.8 &  -- &  LL14-001 &  XMM-001 &  no &  -- &  yes &  yes \\ 
LL14-002 &  01:32:27.85 &  30:35:44.6 &  73 &  6.9 &  -- &  LL14-002 &  XMM-002 &  no &  -- &  yes &  yes \\ 
LL14-004 &  01:32:35.36 &  30:35:19.8 &  86 &  6.2 &  -- &  LL14-004 &  XMM-005 &  no &  -- &  yes &  no \\ 
LL14-005 &  01:32:37.16 &  30:17:54.3 &  86 &  6.8 &  -- &  LL14-005 &  XMM-006 &  yes &  -- &  yes &  no \\ 
LL14-006 &  01:32:39.78 &  30:27:55.0 &  36 &  5.7 &  -- &  LL14-006 &  XMM-007 &  no &  -- &  yes &  yes \\ 
LL14-007 &  01:32:40.23 &  30:16:21.3 &  44 &  6.9 &  -- &  LL14-007 &  XMM-008 &  no &  -- &  yes &  no \\ 
LL14-008 &  01:32:40.68 &  30:16:31.5 &  42 &  6.8 &  -- &  LL14-008 &  XMM-009 &  yes &  -- &  yes &  no \\ 
LL14-009 &  01:32:40.94 &  30:31:51.1 &  87 &  5.6 &  -- &  LL14-009 &  XMM-010 &  no &  -- &  yes &  no \\ 
LL14-010 &  01:32:42.71 &  30:36:20.1 &  59 &  5.6 &  -- &  LL14-010 &  XMM-012 &  no &  -- &  yes &  yes \\ 
LL14-012 &  01:32:45.47 &  30:23:14.2 &  45 &  5.7 &  -- &  LL14-012 &  XMM-014 &  no &  -- &  yes &  no \\ 
LL14-014 &  01:32:51.84 &  30:51:09.0 &  40 &  7.0 &  -- &  LL14-014 &  XMM-016 &  no &  -- &  yes &  no \\ 
LL14-016 &  01:32:52.80 &  30:31:34.2 &  69 &  4.7 &  -- &  LL14-016 &  XMM-018 &  no &  -- &  yes &  yes \\ 
LL14-020 &  01:32:56.12 &  30:33:30.4 &  81 &  4.4 &  -- &  LL14-020 &  XMM-023 &  yes &  -- &  yes &  yes \\ 
LL14-023 &  01:32:57.18 &  30:39:14.7 &  37 &  4.6 &  -- &  LL14-023 &  XMM-025 &  no &  -- &  yes &  no \\ 
LL14-030 &  01:33:08.55 &  30:12:15.2 &  27 &  6.9 &  -- &  LL14-030 &  XMM-034 &  yes &  -- &  yes &  yes \\ 
LL14-032 &  01:33:09.69 &  30:16:39.0 &  86 &  5.8 &  -- &  LL14-032 &  XMM-036 &  no &  -- &  yes &  no \\ 
LL14-038 &  01:33:13.46 &  30:28:13.1 &  75 &  3.5 &  -- &  LL14-038 &  XMM-042 &  yes &  -- &  yes &  yes \\ 
LL14-039 &  01:33:13.81 &  30:39:44.0 &  66 &  3.2 &  -- &  LL14-039 &  XMM-043 &  no &  -- &  no &  - \\ 
LL14-040 &  01:33:15.35 &  30:35:41.9 &  76 &  2.8 &  -- &  LL14-040 &  XMM-044 &  no &  -- &  yes &  yes \\ 
LL14-043 &  01:33:17.55 &  30:46:45.6 &  69 &  4.1 &  -- &  LL14-043 &  XMM-047 &  no &  -- &  yes &  yes \\ 
LL14-044 &  01:33:18.13 &  30:33:38.6 &  30 &  2.7 &  -- &  LL14-044 &  XMM-048 &  no &  -- &  yes &  yes \\ 
LL14-046 &  01:33:19.52 &  30:12:29.3 &  70 &  6.8 &  -- &  LL14-046 &  XMM-051 &  no &  -- &  yes &  no \\ 
LL14-047 &  01:33:20.76 &  30:25:55.2 &  16 &  3.6 &  -- &  LL14-047 &  XMM-052 &  no &  -- &  yes &  yes \\ 
LL14-048 &  01:33:21.19 &  30:19:20.6 &  75 &  5.1 &  -- &  LL14-048 &  XMM-053 &  no &  -- &  yes &  no \\ 
LL14-049 &  01:33:21.33 &  30:30:31.6 &  56 &  2.8 &  -- &  LL14-049 &  XMM-054 &  yes &  -- &  yes &  yes \\ 
LL14-051 &  01:33:21.94 &  30:25:58.4 &  36 &  3.6 &  -- &  LL14-051 &  XMM-056 &  no &  -- &  yes &  yes \\ 
LL14-054 &  01:33:24.01 &  30:36:56.8 &  77 &  2.2 &  -- &  LL14-054 &  XMM-059 &  yes &  -- &  yes &  yes \\ 
LL14-055 &  01:33:24.18 &  30:28:50.2 &  51 &  2.9 &  -- &  LL14-055 &  XMM-060 &  no &  -- &  yes &  yes \\ 
LL14-057 &  01:33:27.32 &  30:23:59.3 &  34 &  3.9 &  -- &  LL14-057 &  XMM-062 &  no &  -- &  yes &  no \\ 
LL14-058 &  01:33:27.92 &  30:18:17.4 &  40 &  5.3 &  -- &  LL14-058 &  XMM-063 &  no &  -- &  yes &  yes \\ 
LL14-059 &  01:33:28.00 &  30:16:01.2 &  37 &  5.9 &  -- &  LL14-059 &  XMM-064 &  no &  -- &  yes &  no \\ 
LL14-063 &  01:33:29.79 &  31:01:53.0 &  48 &  7.0 &  -- &  LL14-063 &  XMM-069 &  yes &  -- &  yes &  yes \\ 
LL14-065 &  01:33:30.64 &  30:21:01.5 &  89 &  4.7 &  -- &  LL14-065 &  XMM-071 &  no &  -- &  yes &  yes \\ 
LL14-066 &  01:33:31.20 &  30:21:14.3 &  59 &  4.6 &  -- &  LL14-066 &  XMM-072 &  no &  -- &  no &  - \\ 
LL14-070 &  01:33:35.10 &  30:19:24.2 &  64 &  5.1 &  -- &  LL14-070 &  XMM-077 &  no &  -- &  yes &  yes \\ 
LL14-072 &  01:33:34.99 &  30:29:54.6 &  18 &  2.5 &  -- &  LL14-072 &  XMM-076 &  yes &  -- &  yes &  yes \\ 
LL14-075 &  01:33:37.02 &  30:33:10.0 &  75 &  1.7 &  -- &  LL14-075 &  XMM-084 &  yes &  -- &  yes &  yes \\ 
LL14-079 &  01:33:38.65 &  31:02:38.8 &  60 &  6.8 &  -- &  LL14-079 &  XMM-087 &  no &  -- &  yes &  yes \\ 
LL14-080 &  01:33:39.59 &  30:34:26.0 &  62 &  1.4 &  -- &  LL14-080 &  XMM-088 &  yes &  -- &  yes &  yes \\ 
LL14-081 &  01:33:40.53 &  30:10:51.7 &  47 &  7.5 &  -- &  LL14-081 &  XMM-090 &  yes &  -- &  yes &  yes \\ 
LL14-092 &  01:33:47.52 &  30:17:13.8 &  57 &  6.0 &  -- &  LL14-092 &  XMM-103 &  no &  -- &  yes &  yes \\ 
LL14-093 &  01:33:47.82 &  30:18:02.1 &  95 &  5.8 &  -- &  LL14-093 &  XMM-104 &  no &  -- &  no &  - \\ 
LL14-094 &  01:33:48.13 &  30:17:25.9 &  19 &  5.9 &  -- &  LL14-094 &  XMM-105 &  no &  -- &  no &  - \\ 
LL14-096 &  01:33:48.92 &  30:33:05.2 &  40 &  1.7 &  -- &  LL14-096 &  -- &  no &  -- &  no &  - \\ 
LL14-103 &  01:33:52.56 &  30:28:38.4 &  40 &  3.0 &  -- &  LL14-103 &  XMM-115 &  no &  -- &  yes &  yes \\ 
LL14-106 &  01:33:54.69 &  30:18:51.0 &  73 &  5.8 &  -- &  LL14-106 &  XMM-117 &  yes &  -- &  yes &  yes \\ 
LL14-109 &  01:33:55.29 &  30:16:49.0 &  31 &  6.4 &  -- &  LL14-109 &  XMM-121 &  no &  -- &  yes &  no \\ 
LL14-114 &  01:33:57.41 &  31:00:55.8 &  61 &  5.6 &  -- &  LL14-114 &  XMM-126 &  no &  -- &  yes &  yes \\ 
LL14-120 &  01:33:59.15 &  30:32:42.1 &  39 &  2.3 &  -- &  LL14-120 &  XMM-133 &  no &  -- &  yes &  yes \\ 
LL14-122 &  01:34:00.25 &  30:39:28.9 &  40 &  0.8 &  -- &  LL14-122 &  XMM-137 &  yes &  -- &  yes &  yes \\ 
LL14-125 &  01:34:00.58 &  30:50:42.9 &  23 &  2.8 &  -- &  LL14-125 &  XMM-138 &  yes &  -- &  yes &  yes \\ 
LL14-128 &  01:34:02.10 &  30:28:34.5 &  37 &  3.5 &  -- &  LL14-128 &  XMM-141 &  no &  -- &  yes &  yes \\ 
LL14-133 &  01:34:04.88 &  30:58:30.7 &  73 &  4.8 &  -- &  LL14-133 &  XMM-146 &  no &  -- &  yes &  no \\ 
LL14-134 &  01:34:05.50 &  31:07:26.4 &  56 &  7.2 &  -- &  LL14-134 &  XMM-147 &  yes &  -- &  yes &  no \\ 
LL14-137 &  01:34:07.98 &  31:01:03.7 &  55 &  5.4 &  -- &  LL14-137 &  XMM-150 &  no &  -- &  yes &  yes \\ 
LL14-142 &  01:34:11.21 &  30:24:15.3 &  42 &  5.1 &  -- &  LL14-142 &  XMM-155 &  no &  -- &  yes &  yes \\ 
LL14-143 &  01:34:12.28 &  31:02:43.4 &  56 &  5.8 &  -- &  LL14-143 &  XMM-156 &  no &  -- &  yes &  no \\ 
LL14-144 &  01:34:12.90 &  30:23:24.3 &  47 &  5.5 &  -- &  LL14-144 &  XMM-158 &  no &  -- &  yes &  no \\ 
LL14-148 &  01:34:13.85 &  30:30:39.8 &  16 &  3.7 &  -- &  LL14-148 &  XMM-162 &  no &  -- &  yes &  no \\ 
LL14-153 &  01:34:14.55 &  30:44:36.2 &  52 &  2.0 &  -- &  LL14-153 &  XMM-167 &  no &  -- &  yes &  yes \\ 
LL14-162 &  01:34:19.45 &  30:52:48.9 &  76 &  3.5 &  -- &  LL14-162 &  XMM-176 &  yes &  -- &  no &  - \\ 
LL14-163 &  01:34:19.68 &  30:33:41.5 &  38 &  3.5 &  -- &  LL14-163 &  XMM-177 &  no &  -- &  no &  - \\ 
LL14-167 &  01:34:24.08 &  30:33:24.4 &  61 &  3.9 &  -- &  LL14-167 &  XMM-181 &  no &  -- &  no &  - \\ 
LL14-168 &  01:34:24.48 &  30:48:58.3 &  81 &  3.0 &  -- &  LL14-168 &  XMM-182 &  no &  -- &  yes &  yes \\ 
LL14-173 &  01:34:29.77 &  30:35:08.4 &  86 &  4.0 &  -- &  LL14-173 &  XMM-187 &  yes &  -- &  yes &  yes \\ 
LL14-174 &  01:34:30.25 &  30:56:36.2 &  34 &  4.5 &  -- &  LL14-174 &  -- &  no &  -- &  no &  - \\ 
LL14-180 &  01:34:37.40 &  30:44:11.0 &  62 &  3.7 &  -- &  LL14-180 &  XMM-194 &  no &  -- &  yes &  yes \\ 
LL14-182 &  01:34:39.69 &  30:39:17.6 &  65 &  4.3 &  -- &  LL14-182 &  XMM-196 &  no &  -- &  yes &  yes \\ 
LL14-183 &  01:34:39.94 &  31:06:02.7 &  54 &  6.8 &  -- &  LL14-183 &  XMM-197 &  no &  -- &  yes &  yes \\ 
LL14-187 &  01:34:42.68 &  30:40:51.5 &  39 &  4.4 &  -- &  LL14-187 &  XMM-202 &  no &  -- &  yes &  yes \\ 
LL14-188 &  01:34:44.02 &  31:01:48.9 &  81 &  6.0 &  -- &  LL14-188 &  XMM-203 &  no &  -- &  yes &  no \\ 
LL14-189 &  01:34:45.40 &  30:35:35.2 &  50 &  5.3 &  -- &  LL14-189 &  XMM-205 &  no &  -- &  yes &  yes \\ 
LL14-190 &  01:34:45.88 &  30:57:19.1 &  51 &  5.3 &  -- &  LL14-190 &  XMM-206 &  no &  -- &  yes &  yes \\ 
LL14-191 &  01:34:47.24 &  30:34:25.0 &  37 &  5.6 &  -- &  LL14-191 &  XMM-207 &  no &  -- &  yes &  no \\ 
LL14-192 &  01:34:50.48 &  31:07:38.6 &  56 &  7.3 &  -- &  LL14-192 &  XMM-208 &  no &  -- &  no &  - \\ 
LL14-193 &  01:34:52.48 &  30:50:22.2 &  85 &  5.0 &  -- &  LL14-193 &  XMM-209 &  no &  -- &  yes &  yes \\ 
LL14-195 &  01:34:58.60 &  31:10:09.4 &  44 &  8.1 &  -- &  LL14-195 &  XMM-212 &  no &  -- &  no &  - \\ 
LL14-196 &  01:34:59.19 &  30:40:16.5 &  36 &  5.9 &  -- &  LL14-196 &  XMM-213 &  no &  -- &  yes &  yes \\ 
LL14-198 &  01:35:00.40 &  31:02:36.3 &  73 &  6.8 &  -- &  LL14-198 &  XMM-215 &  no &  -- &  yes &  no \\ 
LL14-199 &  01:35:01.82 &  30:39:53.9 &  81 &  6.1 &  -- &  LL14-199 &  XMM-217 &  no &  -- &  yes &  yes \\ 
\enddata 
\tablerefs{BK85=\cite{blair85}; G98=\cite{gordon98}; L10=\cite{long10}; S93=\cite{smith93}}
\end{deluxetable}


\clearpage

\startlongtable
\begin{deluxetable}{rrrrrrrrrrc}
\tablecaption{Emission line fluxes of SNR candidates\tablenotemark{a}
\label{table_snr_spec}}
\tablehead{\colhead{Source} & 
 \colhead{H$\alpha$~flux\tablenotemark{b}} & 
 \colhead{H$\beta$} & 
 \colhead{\oiii} & 
 \colhead{\oi} & 
 \colhead{H$\alpha$} & 
 \colhead{\nii} & 
 \colhead{\sii} & 
 \colhead{\sii} & 
 \colhead{\sii:H$\alpha$} & 
 \colhead{FWHM} 
\\
\colhead{~} & 
 \colhead{~} & 
 \colhead{~} & 
 \colhead{$\lambda$5007} & 
 \colhead{$\lambda$6300} & 
 \colhead{~} & 
 \colhead{$\lambda$6584} & 
 \colhead{$\lambda$6717} & 
 \colhead{$\lambda$6731} & 
 \colhead{~} & 
 \colhead{(\AA)} 
}
\tabletypesize{\footnotesize}
\tablewidth{0pt}\startdata
L10-001 &  51 &  95 &  76 &  56 &  300 &  45 &  127 &  90 &  0.72 &  5.7 \\ 
L10-002 &  168 &  95 &  274 &  47 &  300 &  51 &  109 &  77 &  0.62 &  5.0 \\ 
L10-003 &  89 &  92 &  51 &  22 &  300 &  42 &  79 &  57 &  0.45 &  4.9 \\ 
L10-004 &  46 &  88 &  $\sim$36 &  $\sim$62 &  300 &  34 &  129 &  89 &  0.73 &  5.2 \\ 
L10-005 &  32 &  70 &  301 &  $\sim$39 &  300 &  55 &  122 &  83 &  0.69 &  5.3 \\ 
L10-006 &  28 &  116 &  257 &  $\sim$26 &  300 &  64 &  91 &  65 &  0.52 &  5.3 \\ 
L10-007 &  27 &  90 &  34 &  79 &  300 &  35 &  126 &  83 &  0.70 &  5.1 \\ 
L10-008 &  92 &  94 &  30 &  $\sim$25 &  300 &  55 &  112 &  79 &  0.64 &  5.4 \\ 
L10-009 &  120 &  101 &  257 &  44 &  300 &  65 &  154 &  108 &  0.87 &  5.2 \\ 
L10-010 &  55 &  91 &  55 &  50 &  300 &  53 &  125 &  93 &  0.73 &  5.3 \\ 
L10-011 &  828 &  100 &  142 &  77 &  300 &  62 &  136 &  117 &  0.84 &  5.9 \\ 
L10-012 &  377 &  95 &  74 &  18 &  300 &  58 &  110 &  78 &  0.63 &  5.0 \\ 
L10-013 &  29 &  110 &  359 &  -- &  300 &  42 &  94 &  57 &  0.50 &  5.8 \\ 
L10-014 &  339 &  97 &  72 &  26 &  300 &  59 &  114 &  81 &  0.65 &  5.0 \\ 
L10-016 &  65 &  91 &  140 &  48 &  300 &  72 &  125 &  89 &  0.72 &  5.2 \\ 
L10-018 &  252 &  93 &  95 &  41 &  300 &  68 &  115 &  84 &  0.66 &  5.7 \\ 
L10-019 &  105 &  108 &  161 &  27 &  300 &  47 &  99 &  68 &  0.56 &  4.9 \\ 
L10-020 &  33 &  73 &  262 &  $\sim$50 &  300 &  83 &  159 &  115 &  0.91 &  5.0 \\ 
L10-022 &  53 &  89 &  95 &  79 &  300 &  77 &  146 &  104 &  0.83 &  5.1 \\ 
L10-023 &  110 &  88 &  198 &  59 &  300 &  73 &  100 &  86 &  0.62 &  5.7 \\ 
L10-024 &  95 &  55 &  $\sim$21 &  $\sim$18 &  300 &  66 &  91 &  64 &  0.52 &  4.8 \\ 
L10-025 &  313 &  89 &  221 &  $\sim$17 &  300 &  51 &  63 &  61 &  0.41 &  6.0 \\ 
L10-026 &  155 &  94 &  103 &  24 &  300 &  74 &  130 &  92 &  0.74 &  5.0 \\ 
L10-027 &  93 &  81 &  $\sim$27 &  62 &  300 &  76 &  163 &  113 &  0.92 &  5.1 \\ 
L10-028 &  132 &  92 &  69 &  $\sim$14 &  300 &  81 &  84 &  60 &  0.48 &  4.8 \\ 
L10-029 &  65 &  83 &  42 &  88 &  300 &  75 &  196 &  132 &  1.09 &  5.1 \\ 
L10-030 &  129 &  81 &  33 &  $\sim$11 &  300 &  70 &  79 &  57 &  0.45 &  5.0 \\ 
L10-031 &  100 &  90 &  227 &  68 &  300 &  103 &  183 &  136 &  1.06 &  5.8 \\ 
L10-032 &  242 &  72 &  73 &  114 &  300 &  88 &  188 &  139 &  1.09 &  5.8 \\ 
L10-033 &  30 &  $\sim$89 &  -- &  $\sim$81 &  300 &  62 &  167 &  109 &  0.92 &  5.1 \\ 
L10-034 &  105 &  $\sim$51 &  395 &  $\sim$21 &  300 &  101 &  121 &  94 &  0.72 &  6.4 \\ 
L10-035 &  108 &  -- &  -- &  -- &  300 &  -- &  67 &  43 &  0.37 &  5.5 \\ 
L10-036 &  1850 &  87 &  133 &  106 &  300 &  139 &  173 &  168 &  1.14 &  8.3 \\ 
L10-037 &  26 &  85 &  286 &  $\sim$31 &  300 &  78 &  117 &  91 &  0.69 &  6.2 \\ 
L10-038 &  72 &  99 &  84 &  -- &  300 &  69 &  107 &  75 &  0.61 &  4.8 \\ 
L10-039 &  2342 &  83 &  326 &  76 &  300 &  178 &  117 &  155 &  0.91 &  9.1 \\ 
L10-040 &  112 &  97 &  87 &  $\sim$10 &  300 &  63 &  58 &  42 &  0.33 &  5.0 \\ 
L10-041 &  35 &  $\sim$54 &  $\sim$44 &  81 &  300 &  88 &  187 &  138 &  1.08 &  5.0 \\ 
L10-042 &  41 &  $\sim$38 &  58 &  77 &  300 &  60 &  161 &  113 &  0.91 &  5.0 \\ 
L10-043 &  128 &  95 &  44 &  $\sim$20 &  300 &  84 &  105 &  71 &  0.58 &  4.8 \\ 
L10-044 &  68 &  $\sim$78 &  58 &  97 &  300 &  90 &  174 &  122 &  0.99 &  6.0 \\ 
L10-045 &  2485 &  79 &  153 &  39 &  300 &  124 &  131 &  117 &  0.83 &  7.4 \\ 
L10-046 &  77 &  $\sim$89 &  68 &  67 &  300 &  99 &  171 &  126 &  0.99 &  5.7 \\ 
L10-048 &  164 &  $\sim$83 &  20 &  $\sim$12 &  300 &  85 &  97 &  65 &  0.54 &  4.8 \\ 
L10-049 &  86 &  $\sim$64 &  186 &  $\sim$46 &  300 &  134 &  148 &  103 &  0.83 &  5.0 \\ 
L10-050 &  72 &  86 &  54 &  -- &  300 &  96 &  101 &  66 &  0.56 &  5.0 \\ 
L10-051 &  13 &  56 &  $\sim$65 &  -- &  300 &  54 &  146 &  110 &  0.85 &  5.0 \\ 
L10-052 &  106 &  83 &  35 &  -- &  300 &  76 &  90 &  63 &  0.51 &  4.9 \\ 
L10-053 &  220 &  81 &  59 &  28 &  300 &  51 &  102 &  74 &  0.58 &  5.4 \\ 
L10-054 &  114 &  82 &  52 &  36 &  300 &  54 &  129 &  92 &  0.74 &  5.2 \\ 
L10-055 &  168 &  83 &  13 &  $\sim$17 &  300 &  87 &  119 &  85 &  0.68 &  4.8 \\ 
L10-057 &  131 &  79 &  31 &  28 &  300 &  90 &  117 &  81 &  0.66 &  4.9 \\ 
L10-058 &  112 &  93 &  13 &  $\sim$29 &  300 &  86 &  126 &  91 &  0.73 &  4.9 \\ 
L10-059 &  60 &  $\sim$28 &  75 &  $\sim$47 &  300 &  173 &  194 &  140 &  1.11 &  5.2 \\ 
L10-060 &  305 &  72 &  59 &  56 &  300 &  127 &  143 &  102 &  0.82 &  5.2 \\ 
L10-061 &  217 &  85 &  30 &  55 &  300 &  101 &  162 &  114 &  0.92 &  4.9 \\ 
L10-062 &  31 &  67 &  $\sim$21 &  -- &  300 &  90 &  114 &  92 &  0.68 &  4.8 \\ 
L10-063 &  20 &  $\sim$66 &  $\sim$54 &  $\sim$53 &  300 &  101 &  152 &  110 &  0.87 &  4.9 \\ 
L10-064 &  41 &  $\sim$60 &  242 &  -- &  300 &  101 &  129 &  90 &  0.73 &  5.5 \\ 
L10-065 &  146 &  82 &  32 &  $\sim$25 &  300 &  108 &  119 &  83 &  0.67 &  4.9 \\ 
L10-066 &  21 &  64 &  $\sim$52 &  $\sim$86 &  300 &  111 &  207 &  145 &  1.17 &  5.1 \\ 
L10-067 &  61 &  81 &  69 &  $\sim$49 &  300 &  91 &  143 &  101 &  0.81 &  5.1 \\ 
L10-068 &  52 &  97 &  294 &  60 &  300 &  87 &  158 &  107 &  0.89 &  4.8 \\ 
L10-069 &  142 &  $\sim$71 &  38 &  $\sim$33 &  300 &  88 &  142 &  101 &  0.81 &  5.1 \\ 
L10-070 &  416 &  90 &  74 &  57 &  300 &  117 &  164 &  119 &  0.94 &  5.3 \\ 
L10-071 &  1113 &  94 &  194 &  86 &  300 &  130 &  172 &  155 &  1.09 &  7.4 \\ 
L10-072 &  54 &  $\sim$31 &  93 &  $\sim$60 &  300 &  190 &  206 &  152 &  1.19 &  4.9 \\ 
L10-073 &  65 &  94 &  119 &  $\sim$35 &  300 &  70 &  144 &  97 &  0.80 &  5.1 \\ 
L10-074 &  151 &  79 &  49 &  87 &  300 &  102 &  191 &  133 &  1.08 &  5.7 \\ 
L10-075 &  43 &  $\sim$47 &  $\sim$26 &  74 &  300 &  132 &  206 &  133 &  1.13 &  4.9 \\ 
L10-076 &  180 &  90 &  107 &  34 &  300 &  114 &  124 &  87 &  0.70 &  4.8 \\ 
L10-077 &  258 &  88 &  17 &  $\sim$8 &  300 &  67 &  78 &  53 &  0.44 &  4.8 \\ 
L10-078 &  295 &  88 &  160 &  93 &  300 &  161 &  206 &  152 &  1.19 &  5.5 \\ 
L10-079 &  148 &  90 &  14 &  $\sim$20 &  300 &  74 &  102 &  70 &  0.57 &  4.9 \\ 
L10-080 &  175 &  62 &  123 &  95 &  300 &  159 &  198 &  159 &  1.19 &  5.5 \\ 
L10-081 &  77 &  $\sim$66 &  164 &  -- &  300 &  89 &  111 &  82 &  0.64 &  4.8 \\ 
L10-082 &  33 &  $\sim$65 &  303 &  $\sim$41 &  300 &  95 &  180 &  123 &  1.01 &  6.2 \\ 
L10-084 &  55 &  $\sim$61 &  409 &  $\sim$39 &  300 &  178 &  214 &  148 &  1.20 &  9.0 \\ 
L10-085 &  138 &  73 &  223 &  82 &  300 &  136 &  188 &  146 &  1.11 &  5.3 \\ 
L10-086 &  48 &  71 &  85 &  $\sim$69 &  300 &  90 &  95 &  96 &  0.64 &  5.1 \\ 
L10-087 &  41 &  65 &  75 &  -- &  300 &  87 &  137 &  106 &  0.81 &  5.0 \\ 
L10-088 &  30 &  67 &  113 &  $\sim$59 &  300 &  102 &  170 &  131 &  1.00 &  5.1 \\ 
L10-089 &  144 &  83 &  35 &  $\sim$18 &  300 &  84 &  102 &  74 &  0.59 &  4.8 \\ 
L10-090 &  19 &  $\sim$51 &  505 &  -- &  300 &  139 &  191 &  133 &  1.08 &  5.9 \\ 
L10-091 &  27 &  86 &  346 &  $\sim$48 &  300 &  126 &  199 &  138 &  1.12 &  5.7 \\ 
L10-092 &  79 &  81 &  100 &  42 &  300 &  106 &  143 &  99 &  0.81 &  4.9 \\ 
L10-093 &  21 &  $\sim$19 &  173 &  $\sim$68 &  300 &  103 &  174 &  121 &  0.98 &  4.9 \\ 
L10-094 &  96 &  $\sim$75 &  344 &  $\sim$14 &  300 &  110 &  131 &  102 &  0.78 &  5.1 \\ 
L10-096 &  987 &  90 &  51 &  148 &  300 &  118 &  235 &  195 &  1.43 &  5.5 \\ 
L10-097 &  96 &  80 &  159 &  61 &  300 &  118 &  153 &  117 &  0.90 &  5.6 \\ 
L10-098 &  94 &  88 &  106 &  $\sim$15 &  300 &  62 &  76 &  53 &  0.43 &  4.8 \\ 
L10-099 &  155 &  85 &  30 &  27 &  300 &  90 &  145 &  105 &  0.83 &  4.9 \\ 
L10-100 &  37 &  69 &  292 &  -- &  300 &  98 &  152 &  106 &  0.86 &  5.8 \\ 
L10-101 &  240 &  94 &  18 &  25 &  300 &  90 &  151 &  105 &  0.85 &  4.9 \\ 
L10-103 &  26 &  62 &  182 &  -- &  300 &  119 &  160 &  111 &  0.90 &  5.4 \\ 
L10-104 &  45 &  $\sim$73 &  46 &  $\sim$80 &  300 &  101 &  172 &  132 &  1.01 &  5.5 \\ 
L10-105 &  57 &  82 &  131 &  62 &  300 &  83 &  171 &  121 &  0.97 &  5.8 \\ 
L10-106 &  17 &  $\sim$70 &  249 &  $\sim$91 &  300 &  104 &  175 &  119 &  0.98 &  4.8 \\ 
L10-107 &  66 &  85 &  247 &  $\sim$34 &  300 &  93 &  141 &  100 &  0.80 &  6.0 \\ 
L10-108 &  60 &  95 &  98 &  $\sim$14 &  300 &  61 &  95 &  71 &  0.55 &  4.9 \\ 
L10-109 &  389 &  92 &  225 &  10 &  300 &  60 &  68 &  48 &  0.39 &  5.0 \\ 
L10-110 &  151 &  96 &  43 &  26 &  300 &  59 &  101 &  71 &  0.57 &  5.0 \\ 
L10-111 &  55 &  74 &  183 &  $\sim$47 &  300 &  120 &  172 &  116 &  0.96 &  7.7 \\ 
L10-112 &  19 &  $\sim$43 &  51 &  -- &  300 &  100 &  191 &  144 &  1.11 &  4.5 \\ 
L10-113 &  183 &  85 &  284 &  $\sim$15 &  300 &  55 &  82 &  57 &  0.46 &  5.0 \\ 
L10-114 &  391 &  98 &  316 &  40 &  300 &  72 &  124 &  89 &  0.71 &  5.1 \\ 
L10-115 &  38 &  99 &  90 &  69 &  300 &  62 &  140 &  96 &  0.79 &  5.9 \\ 
L10-116 &  18 &  $\sim$71 &  239 &  -- &  300 &  93 &  162 &  113 &  0.92 &  4.7 \\ 
L10-117 &  45 &  91 &  81 &  $\sim$58 &  300 &  73 &  146 &  100 &  0.82 &  6.1 \\ 
L10-118 &  30 &  $\sim$70 &  $\sim$67 &  -- &  300 &  64 &  122 &  94 &  0.72 &  5.5 \\ 
L10-119 &  18 &  81 &  $\sim$54 &  -- &  300 &  63 &  81 &  57 &  0.46 &  4.9 \\ 
L10-120 &  113 &  100 &  376 &  22 &  300 &  97 &  102 &  70 &  0.57 &  4.9 \\ 
L10-121 &  28 &  70 &  295 &  -- &  300 &  52 &  75 &  46 &  0.40 &  7.1 \\ 
L10-122 &  102 &  94 &  275 &  37 &  300 &  80 &  162 &  118 &  0.93 &  5.1 \\ 
L10-123 &  56 &  90 &  86 &  106 &  300 &  77 &  188 &  132 &  1.06 &  5.6 \\ 
L10-124 &  1527 &  92 &  112 &  42 &  300 &  68 &  107 &  86 &  0.64 &  5.4 \\ 
L10-125 &  80 &  97 &  280 &  54 &  300 &  102 &  165 &  118 &  0.94 &  5.3 \\ 
L10-126 &  62 &  100 &  693 &  $\sim$49 &  300 &  110 &  106 &  71 &  0.59 &  4.8 \\ 
L10-127 &  61 &  87 &  -- &  $\sim$24 &  300 &  52 &  102 &  73 &  0.58 &  4.9 \\ 
L10-128 &  233 &  97 &  151 &  32 &  300 &  82 &  131 &  90 &  0.74 &  5.7 \\ 
L10-130 &  85 &  95 &  203 &  $\sim$37 &  300 &  85 &  152 &  103 &  0.85 &  5.3 \\ 
L10-131 &  38 &  115 &  254 &  -- &  300 &  52 &  73 &  49 &  0.41 &  4.9 \\ 
L10-132 &  134 &  87 &  14 &  -- &  300 &  40 &  45 &  32 &  0.26 &  4.8 \\ 
L10-133 &  116 &  96 &  33 &  $\sim$13 &  300 &  52 &  96 &  67 &  0.54 &  4.9 \\ 
L10-134 &  18 &  100 &  153 &  $\sim$58 &  300 &  57 &  125 &  80 &  0.68 &  5.0 \\ 
L10-135 &  32 &  95 &  $\sim$55 &  -- &  300 &  63 &  119 &  83 &  0.67 &  5.0 \\ 
L10-136 &  53 &  107 &  45 &  -- &  300 &  52 &  63 &  44 &  0.36 &  4.9 \\ 
L10-137 &  40 &  95 &  $\sim$40 &  $\sim$39 &  300 &  34 &  126 &  83 &  0.69 &  5.1 \\ 
LL14-001 &  18 &  113 &  -- &  -- &  300 &  29 &  95 &  78 &  0.58 &  4.8 \\ 
LL14-002 &  17 &  92 &  138 &  $\sim$49 &  300 &  35 &  115 &  79 &  0.65 &  5.2 \\ 
LL14-004 &  37 &  104 &  59 &  -- &  300 &  30 &  $\sim$37 &  $\sim$22 &  $\sim$0.20 &  4.8 \\ 
LL14-005 &  8 &  119 &  621 &  -- &  300 &  36 &  65 &  47 &  0.37 &  4.6 \\ 
LL14-006 &  15 &  $\sim$79 &  -- &  $\sim$48 &  300 &  50 &  109 &  74 &  0.61 &  5.0 \\ 
LL14-007 &  19 &  90 &  -- &  -- &  300 &  30 &  $\sim$53 &  $\sim$22 &  $\sim$0.25 &  5.0 \\ 
LL14-008 &  10 &  $\sim$76 &  -- &  -- &  300 &  39 &  $\sim$51 &  $\sim$30 &  $\sim$0.27 &  4.6 \\ 
LL14-009 &  23 &  98 &  114 &  -- &  300 &  29 &  41 &  27 &  0.22 &  4.7 \\ 
LL14-010 &  19 &  $\sim$77 &  -- &  -- &  300 &  39 &  88 &  53 &  0.47 &  5.1 \\ 
LL14-012 &  24 &  107 &  -- &  -- &  300 &  43 &  67 &  44 &  0.37 &  4.8 \\ 
LL14-014 &  4 &  $\sim$83 &  402 &  -- &  300 &  -8 &  $\sim$89 &  $\sim$30 &  $\sim$0.40 &  4.6 \\ 
LL14-016 &  35 &  88 &  -- &  -- &  300 &  44 &  74 &  56 &  0.43 &  4.9 \\ 
LL14-020 &  23 &  96 &  83 &  $\sim$50 &  300 &  73 &  153 &  107 &  0.87 &  5.1 \\ 
LL14-023 &  44 &  90 &  $\sim$29 &  $\sim$33 &  300 &  40 &  66 &  43 &  0.36 &  4.9 \\ 
LL14-030 &  19 &  108 &  415 &  -- &  300 &  35 &  81 &  64 &  0.48 &  5.9 \\ 
LL14-032 &  6 &  -- &  -- &  -- &  300 &  -- &  -- &  -- &  -- &  5.5 \\ 
LL14-038 &  17 &  86 &  136 &  -- &  300 &  82 &  126 &  93 &  0.73 &  5.1 \\ 
LL14-040 &  8 &  -- &  -- &  $\sim$151 &  300 &  84 &  133 &  90 &  0.74 &  4.9 \\ 
LL14-043 &  21 &  88 &  117 &  -- &  300 &  63 &  128 &  87 &  0.72 &  5.0 \\ 
LL14-044 &  26 &  44 &  -- &  -- &  300 &  54 &  78 &  52 &  0.43 &  4.6 \\ 
LL14-046 &  5 &  $\sim$126 &  -- &  -- &  300 &  12 &  $\sim$82 &  $\sim$38 &  $\sim$0.40 &  4.5 \\ 
LL14-047 &  48 &  82 &  427 &  -- &  300 &  51 &  69 &  52 &  0.40 &  4.7 \\ 
LL14-048 &  18 &  -- &  -- &  -- &  300 &  $\sim$16 &  $\sim$35 &  $\sim$33 &  $\sim$0.23 &  6.2 \\ 
LL14-049 &  56 &  91 &  $\sim$46 &  -- &  300 &  54 &  73 &  49 &  0.41 &  4.9 \\ 
LL14-051 &  60 &  85 &  -- &  -- &  300 &  51 &  92 &  65 &  0.52 &  4.8 \\ 
LL14-054 &  18 &  $\sim$58 &  -- &  -- &  300 &  78 &  102 &  58 &  0.53 &  4.6 \\ 
LL14-055 &  4 &  -- &  $\sim$484 &  -- &  300 &  $\sim$136 &  253 &  173 &  1.42 &  4.7 \\ 
LL14-057 &  28 &  59 &  $\sim$18 &  -- &  300 &  49 &  57 &  38 &  0.31 &  4.6 \\ 
LL14-058 &  15 &  70 &  405 &  $\sim$93 &  300 &  64 &  131 &  108 &  0.80 &  5.9 \\ 
LL14-059 &  9 &  97 &  -- &  -- &  300 &  $\sim$40 &  46 &  64 &  0.37 &  4.7 \\ 
LL14-063 &  2 &  -- &  $\sim$288 &  -- &  300 &  -- &  $\sim$102 &  $\sim$146 &  $\sim$0.83 &  6.2 \\ 
LL14-065 &  11 &  $\sim$71 &  -- &  -- &  300 &  $\sim$45 &  149 &  91 &  0.80 &  4.6 \\ 
LL14-070 &  10 &  $\sim$29 &  -- &  $\sim$124 &  300 &  35 &  $\sim$116 &  $\sim$56 &  $\sim$0.58 &  5.0 \\ 
LL14-072 &  250 &  90 &  239 &  55 &  300 &  96 &  170 &  126 &  0.99 &  5.8 \\ 
LL14-075 &  68 &  77 &  70 &  -- &  300 &  81 &  123 &  70 &  0.64 &  5.1 \\ 
LL14-079 &  21 &  78 &  98 &  $\sim$80 &  300 &  51 &  135 &  93 &  0.76 &  6.9 \\ 
LL14-080 &  37 &  -- &  109 &  -- &  300 &  96 &  130 &  93 &  0.75 &  4.7 \\ 
LL14-081 &  33 &  82 &  68 &  79 &  300 &  37 &  107 &  76 &  0.61 &  5.1 \\ 
LL14-092 &  11 &  $\sim$38 &  501 &  -- &  300 &  40 &  $\sim$82 &  $\sim$46 &  $\sim$0.43 &  4.5 \\ 
LL14-103 &  25 &  72 &  -- &  -- &  300 &  65 &  86 &  62 &  0.49 &  4.9 \\ 
LL14-106 &  17 &  90 &  104 &  -- &  300 &  52 &  122 &  85 &  0.69 &  6.3 \\ 
LL14-109 &  13 &  88 &  -- &  -- &  300 &  40 &  43 &  30 &  0.25 &  4.9 \\ 
LL14-114 &  14 &  97 &  -- &  -- &  300 &  13 &  83 &  47 &  0.43 &  5.0 \\ 
LL14-120 &  87 &  89 &  69 &  -- &  300 &  81 &  110 &  76 &  0.62 &  4.8 \\ 
LL14-122 &  50 &  $\sim$41 &  91 &  -- &  300 &  136 &  121 &  86 &  0.69 &  4.7 \\ 
LL14-125 &  20 &  66 &  184 &  -- &  300 &  95 &  179 &  119 &  0.99 &  5.1 \\ 
LL14-128 &  9 &  $\sim$55 &  -- &  -- &  300 &  63 &  111 &  90 &  0.67 &  4.6 \\ 
LL14-133 &  16 &  71 &  $\sim$54 &  -- &  300 &  14 &  28 &  19 &  0.15 &  4.8 \\ 
LL14-134 &  5 &  $\sim$79 &  $\sim$177 &  -- &  300 &  -- &  -- &  -- &  -- &  4.2 \\ 
LL14-137 &  22 &  104 &  -- &  -- &  300 &  42 &  91 &  61 &  0.50 &  5.7 \\ 
LL14-142 &  11 &  85 &  -- &  -- &  300 &  50 &  81 &  53 &  0.45 &  4.7 \\ 
LL14-143 &  12 &  -- &  -- &  -- &  300 &  $\sim$23 &  46 &  29 &  0.25 &  5.2 \\ 
LL14-144 &  13 &  $\sim$113 &  -- &  -- &  300 &  $\sim$34 &  $\sim$63 &  $\sim$45 &  $\sim$0.36 &  4.7 \\ 
LL14-148 &  59 &  82 &  495 &  -- &  300 &  58 &  59 &  45 &  0.35 &  4.8 \\ 
LL14-153 &  22 &  55 &  106 &  -- &  300 &  89 &  112 &  78 &  0.63 &  5.1 \\ 
LL14-168 &  22 &  63 &  $\sim$33 &  -- &  300 &  62 &  87 &  64 &  0.50 &  4.9 \\ 
LL14-173 &  7 &  $\sim$59 &  -- &  -- &  300 &  $\sim$78 &  $\sim$96 &  $\sim$79 &  $\sim$0.58 &  4.5 \\ 
LL14-180 &  37 &  84 &  37 &  -- &  300 &  52 &  79 &  55 &  0.44 &  4.9 \\ 
LL14-182 &  13 &  63 &  -- &  -- &  300 &  $\sim$31 &  103 &  64 &  0.56 &  5.2 \\ 
LL14-183 &  11 &  88 &  -- &  -- &  300 &  36 &  113 &  78 &  0.64 &  6.0 \\ 
LL14-187 &  27 &  88 &  -- &  -- &  300 &  42 &  75 &  51 &  0.42 &  4.9 \\ 
LL14-188 &  8 &  $\sim$71 &  -- &  -- &  300 &  -- &  -- &  -- &  -- &  4.7 \\ 
LL14-189 &  8 &  88 &  218 &  -- &  300 &  70 &  175 &  95 &  0.90 &  5.1 \\ 
LL14-190 &  3 &  -- &  547 &  -- &  300 &  -- &  -- &  -- &  -- &  5.3 \\ 
LL14-191 &  21 &  79 &  -- &  $\sim$71 &  300 &  49 &  58 &  38 &  0.32 &  5.0 \\ 
LL14-193 &  3 &  -- &  $\sim$262 &  -- &  300 &  -- &  165 &  165 &  1.10 &  4.2 \\ 
LL14-196 &  31 &  103 &  134 &  49 &  300 &  53 &  132 &  93 &  0.75 &  5.3 \\ 
LL14-198 &  10 &  -- &  $\sim$83 &  -- &  300 &  $\sim$22 &  $\sim$38 &  $\sim$20 &  $\sim$0.19 &  4.6 \\ 
LL14-199 &  11 &  116 &  -- &  -- &  300 &  42 &  105 &  77 &  0.61 &  5.1 \\ 
\enddata 
\tablenotetext{a}{ Emission line strengths are listed relative to H$\alpha$ set to 300.}
\tablenotetext{b}{ H$\alpha$ Flux is in ergs cm$^{-2}$ s$^{-1}$ in the 1.5\arcsec\ diameter fiber of HectoSpec.}

\end{deluxetable}



%

\clearpage
\pagebreak

\vspace{5mm}

\end{document}